\documentclass[prd,twocolumn,showpacs,superscriptaddress,nofootinbib]{revtex4-1}

\usepackage{graphicx}
\usepackage{mathtools}
\usepackage{color}
\usepackage{multirow}

\def\healpix{ {\sc HEALPix}}
\def\lapack{ {\sc LAPACK}}
\def\camb{{\sc CAMB}}

\def\pthree{{\bf Set-3}}
\def\pseven{{\bf Set-7}}

\begin{document}

\title{Measuring gravitational lensing of the cosmic microwave background \\using cross correlation with large scale structure}

\author{Chang Feng}
\affiliation{Center for Astrophysics and Space Sciences and the Ax
Center for Experimental Cosmology, University of California San
Diego, La Jolla, CA 92093}

\author{Grigor Aslanyan}
\affiliation{Department of Physics, University of California at San
Diego, La Jolla, CA 92093}

\author{Aneesh V.~Manohar}
\affiliation{Department of Physics, University of California at San
Diego, La Jolla, CA 92093}

\author{Brian Keating}
\affiliation{Center for Astrophysics and Space Sciences and the Ax
Center for Experimental Cosmology, University of California San
Diego, La Jolla, CA 92093}
\affiliation{Department of Physics, University of California at San
Diego, La Jolla, CA 92093}

\author{Hans P. Paar}
\affiliation{Center for Astrophysics and
Space Sciences and the Ax Center for Experimental Cosmology,
University of California San Diego, La Jolla, CA 92093}
\affiliation{Department of Physics, University of California at San
Diego, La Jolla, CA 92093}

\author{Oliver Zahn}
\affiliation{Berkeley Center for Cosmological Physics and Lawrence
Berkeley Laboratory, University of California, Berkeley, CA 94720}

\begin{abstract}
We cross correlate the gravitational lensing map extracted from cosmic microwave background measurements by the Wilkinson Microwave Anisotropy Probe (WMAP) with the radio galaxy distribution from the NRAO VLA Sky Survey (NVSS) by using a quadratic estimator technique. We use the full covariance matrix to filter the data,  and calculate the cross-power spectra for the lensing-galaxy correlation. We explore the impact of changing the values of cosmological parameters on the lensing reconstruction, and obtain statistical detection significances at $>3\sigma$. The results of all cross correlations pass the curl null test as well as a complementary diagnostic test using the NVSS data in equatorial coordinates. We forecast the potential for Planck and NVSS to constrain the lensing-galaxy cross correlation as well as the galaxy bias. The lensing-galaxy cross-power spectra are found to be Gaussian distributed.
\end{abstract}

\pacs{98.70.Vc, 98.62.Sb, 98.80.Es}

\maketitle

\section{introduction}

The cosmic microwave background (CMB) temperature anisotropy contains a wealth of cosmological information and has played a pivotal role in our understanding of the Universe. Besides the primordial fluctuations, various secondary anisotropies,  e.g.\ gravitational lensing, the thermal Sunyaev-Zel'dovich effect,  the kinetic Sunyaev-Zel'dovich effect, as well as the integrated Sachs-Wolfe effect, are playing an increasingly important role in constraining cosmological constituents and dynamics.

Among the secondary effects imprinted on the CMB gravitational lensing is of great importance. The projected gravitational lensing potential is a line-of-sight probe which contains information about the geometric distance traversed by CMB photons and time-dependent gravitational potentials. As such it is very sensitive to  late universe parameters, such as the sum of neutrino masses, the dark energy equation of state and spatial curvature. Since the projected gravitational lensing potential contains both geometric and structure growth information, it effectively breaks the angular diameter distance degeneracy~\cite{Smith:2008an}. Gravitational lensing measurements can also be used to de-lens the $B$-mode polarization of the CMB~\cite{Smith:2010gu}, enabling us to learn about primordial gravitational waves~\cite{Kamionkowski:1996zd} and the energy scale of inflation.

Tentative CMB weak lensing searches have been done with WMAP-7 year data sets~\cite{Smidt:2010by,Feng:2011jx} using non-Gaussian statistics. However, WMAP-7 alone cannot detect weak lensing of the CMB because WMAP temperature maps have insufficient sensitivity~\cite{Feng:2011jx}. Recently, the Atacama Cosmology Telescope ~\cite{Das:2011ak} and South Pole Telescope (SPT)~\cite{vanEngelen:2012va} have performed the first internal lensing reconstruction detections using non-Gaussianity. In addition, Atacama Cosmology Telescope and SPT also measured the gravitational lensing signal from the smoothing effects of the acoustic peaks on the CMB temperature power spectrum~\cite{Reichardt:2008ay,Das:2010ga,Keisler:2011aw}. As the experimental sensitivity improves, internal measurements, either from the power spectrum or the trispectrum, will become more precise in the near future.

The correlation between lensing and large scale structure arises from large scale structure, which deflects CMB photons in the late universe. The signal-to-noise ratio of lensing measurements can be enhanced if the CMB maps are cross correlated with highly sensitive large scale structure tracers, such as luminous red galaxies (LRGs) (which cover the redshift range $0.2<z<0.7$), quasars (which covers the redshift range $z<2.7$) from the Sloan Digital Sky Survey (SDSS),  or the NVSS of radio galaxies which has a higher mean redshift ($z\sim1$) than the LRGs and quasars. Hirata \textit{et al}.~\cite{Hirata:2004rp} used the cross correlation between WMAP-1 and LRGs and quasars from SDSS imaging and found no statistically significant signal. Then Smith \textit{et al}.~\cite{Smith:2007rg} used the cross correlation between WMAP-3 and NVSS, and found a $3.4\sigma$ signal, including systematics. Using a slightly less optimal estimator than Ref.~\cite{Smith:2007rg}, Hirata \textit{et al}.~\cite{Hirata:2008cb} obtained results consistent with, though at slightly lower significance than, Ref.~\cite{Smith:2007rg} for WMAP-3 with LRGs (0.95$\sigma$), WMAP-3 with quasars (1.64$\sigma$), and WMAP-3 with NVSS (2.13$\sigma$) respectively. Recently, SPT found a greater than $4\sigma$ cross correlation between the SPT convergence field and the galaxy survey from the Blanco Cosmology Survey, the Wide-field Infrared Survey Explorer, and Spitzer~\cite{Bleem:2012gm}. In this work, we use WMAP data released in years 1, 3, 5, 7, with NVSS to probe the lensing-galaxy correlation. We follow the methods developed in  Smith \textit{et al}.~\cite{Smith:2007rg} and Hirata \textit{et al}.~\cite{Hirata:2008cb}  using all of WMAP's datasets and compare our results to these earlier analyses.

The structure of the paper is as follows. We introduce the data sets in Sec.~\ref{sec:data}. Gravitational lensing effects on the CMB as well as the lensing extraction technique  are reviewed in Sec.~\ref{sec:lensing}. We describe the cross correlation estimators in Sec.~\ref{sec:cross}, and the forecast for Planck in Sec.~\ref{sec:forcast}. We discuss our results in Sec.~\ref{sec:con}.

\begin{figure*}
\includegraphics[width=6cm]{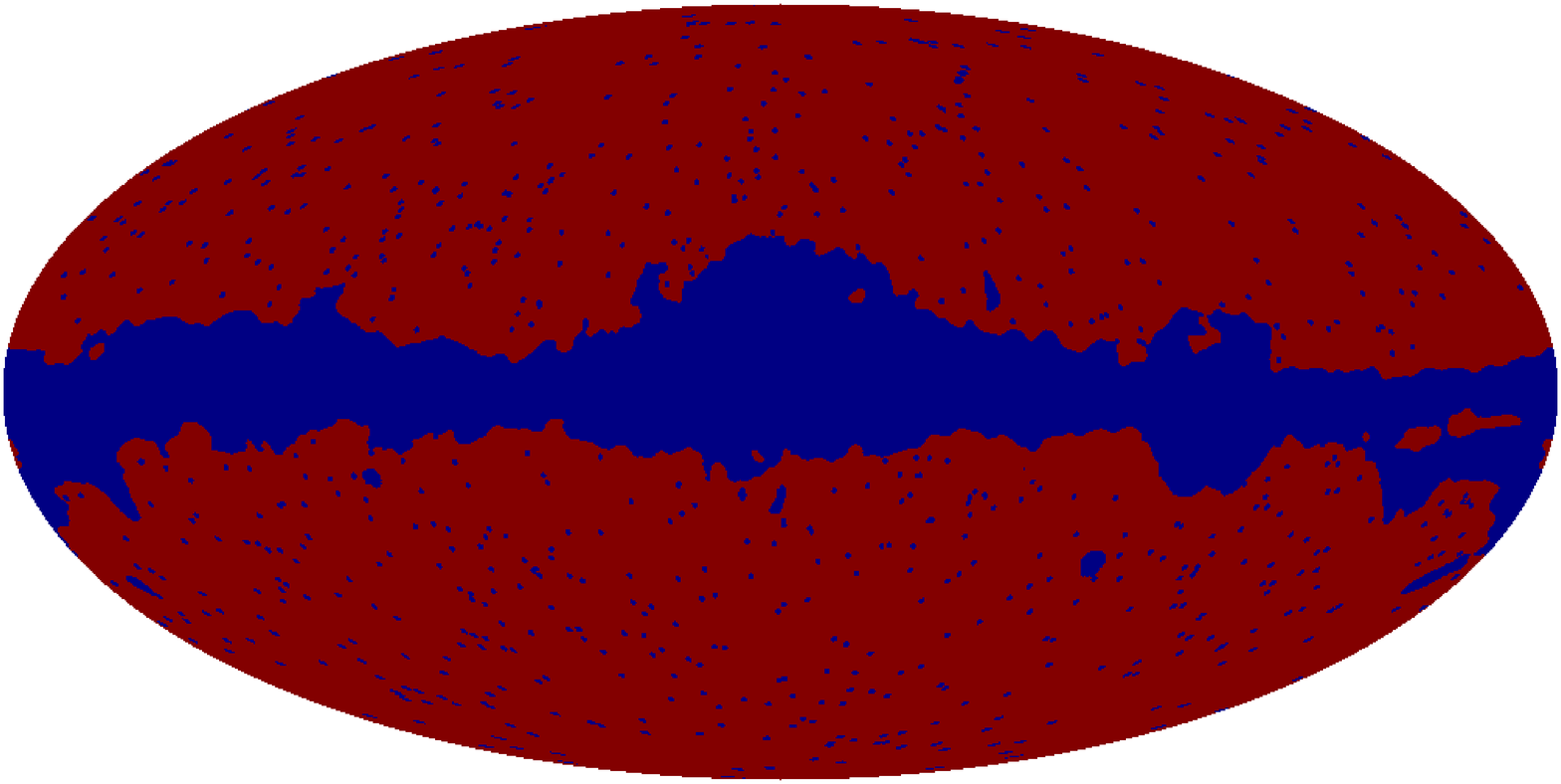}\qquad
\includegraphics[width=6cm]{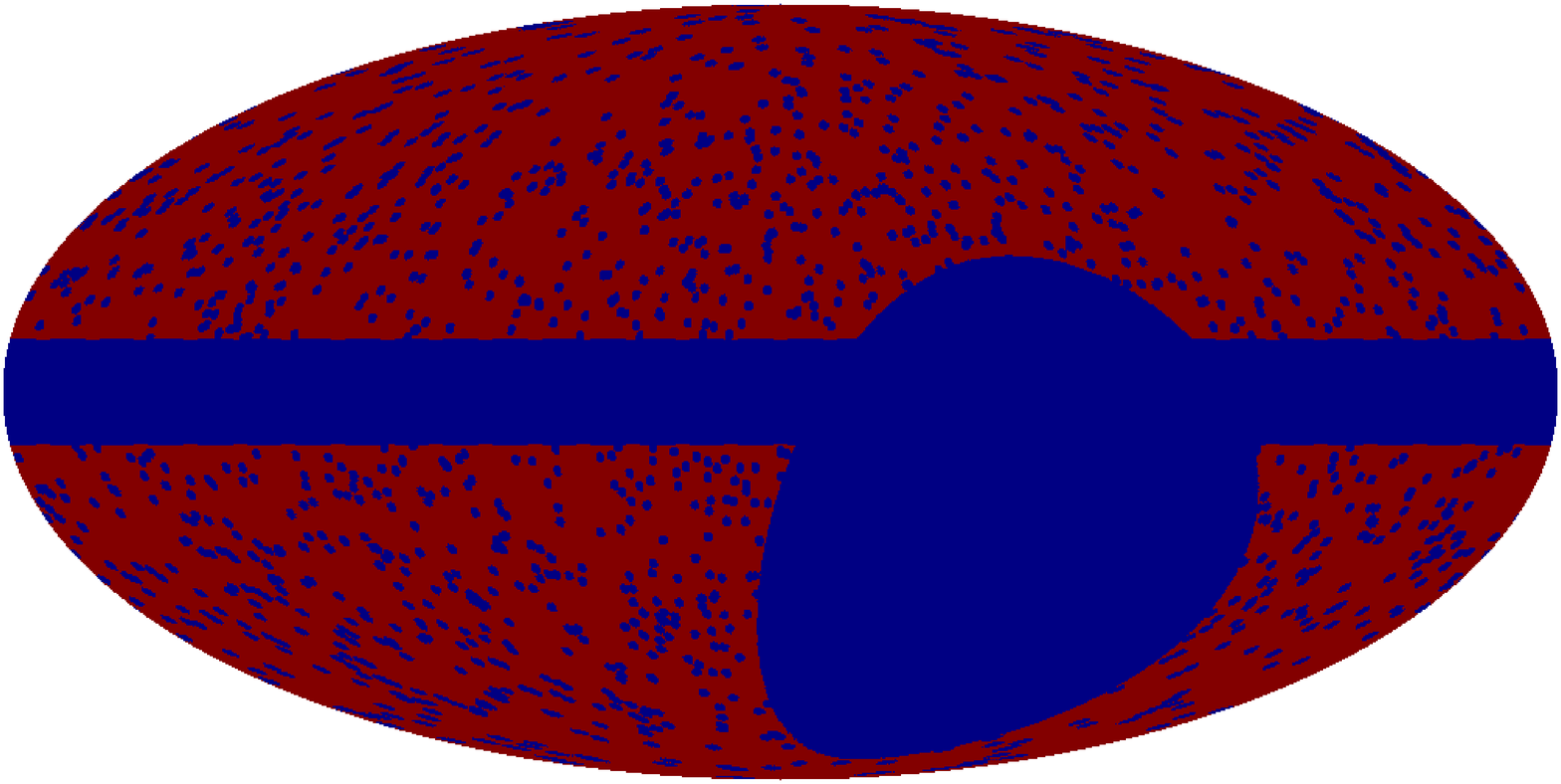}
\caption{WMAP Kp0 mask (left) with $f_{\rm sky}=0.77$ and NVSS mask (right) with $f_{\rm sky}=0.573$.}\label{mask}
\end{figure*}
\section{Data Sets:  WMAP and NVSS}\label{sec:data} 

The CMB data we use are from WMAP's Q-, V-, and W-band raw differencing assemblies (DAs). All of these DAs are masked by the Kp0 mask (Fig.~\ref{mask}) to remove bright sources and the galactic plane leaving $77\%$ of the sky.

The input for the galaxy distribution is the NVSS of radio galaxies. The NVSS~\cite{Condon:1998iy} team provides the software ``NVSSlist" to convert its raw catalog to a deconvolved one which is corrected for known biases and systematic errors. We use the deconvolved catalog to extract the galaxy count map. We use this software here, without specifying  either a minimum or maximum flux cut. The NVSS map is pixelized with a \healpix\ pixelization scheme with $N_{\rm side}=256$. We remove the galactic plane ($|b|<10^\circ$) and the part of the sky unobserved by the survey ($\delta<-36.87^\circ$).  We also carefully remove bright sources with flux $> 1\,\rm Jy$ and mask out a disk of radius $1^{\circ}$ around them, forming the NVSS mask shown in Fig.~\ref{mask}. The resulting galaxy count map has $1,224,990$ sources, a sky fraction $f_{\rm sky}=0.573$, a mean number of sources per pixel $\bar n=2.72$, and a surface density of 170,249 galaxies per steradian. This agrees well with previous studies~\cite{Liu:2010re}. 

\section{Gravitational lensing of the CMB}
\label{sec:lensing}

The effect of lensing on the CMB's primordial temperature $\tilde{T}$ in direction ${\bf n}$ can be represented by
\begin{equation}
T(\mathbf{n})=\widetilde{T}(\mathbf{n}+\mathbf{d}(\mathbf{n})),
\label{cmblensing}
\end{equation}
where $T$ is the lensed temperature and the deflection angle field $\mathbf{d}(\mathbf{n}) = \mathbf{\nabla} \phi$, with $\phi$ being the lensing potential. The operator $\nabla$ is the covariance derivative on the sphere with respect to
the angular position ${\bf n}$. We use Gaussian natural units with $\hbar=c=1$ throughout this paper.

The two-point correlation function of the temperature field is~\cite{Okamoto:2003zw}:
\begin{eqnarray}
\left\langle T_{lm} T_{l'm'} \right\rangle = \tilde{C}^{TT}_l\delta_{ll'}\delta_{m-m'} (-1)^m + \hspace{15mm} \nonumber \\
\sum_{LM}(-1)^M \left(
\begin{array}{ccc}
l   &  l'  & L\\
m & m'&-M
\end{array} \right) f^{TT}_{lLl'}\ \phi_{LM}, \label{lensingcouplemode}
\end{eqnarray}
where the second term encodes the effects of lensing with the weighting factor $f^{TT}_{lLl'}$ given by
\begin{eqnarray}
f^{TT}_{lLl'} = \tilde{C}_l^{TT} {} _0F_{l'Ll} + \tilde{C}_{l'}^{TT}
{} _0F_{lLl'}. \label{weighting}
\end{eqnarray}
We use the standard spherical harmonic decomposition
\begin{equation}
T({\bf n})=\displaystyle\sum_{lm}T_{lm}Y_{lm}({\bf n}),
\end{equation}
which defines the temperature modes $T_{lm}$. We use a similar notation for all other quantities defined on a sphere, e.g.\ 
 $\phi_{LM}$ are the modes of the lensing potential, etc.
 
Here $\tilde{C}_l^{TT}$ is the unlensed temperature power spectrum, and
\begin{eqnarray}
_0F_{lLl'} = \sqrt{\frac{(2l+1)(2l'+1)(2L+1)}{4\pi}} \times \hspace{15mm} \nonumber\\
\frac{1}{2} [L(L+1)+l'(l'+1)-l(l+1)] \left(
\begin{array}{ccc}
l  &  L &l'\\
0 & 0 & 0
\end{array} \right)
\end{eqnarray}
are proportional to the Wigner $3j$ symbols. Equation~(\ref{lensingcouplemode}) provides a way to extract $\phi_{LM}$ from the $TT$ correlations.

In the late universe, the Poisson equation relates the lensing potential $\Phi(\mathbf{k})$ to the density contrast $\delta(\mathbf{k})$,
\begin{equation}
k^2\Phi(\mathbf{k})=\frac{3H_0^2\Omega_m}{2a}\delta(\mathbf{k}),
\label{possion}
\end{equation}
where $\Omega_m$ is the matter fraction, $a$ is the scale factor and $H_0$ is the Hubble constant. Using the definition $D(\chi)=-2(1/\chi-1/\chi_*)$,  the projected lensing potential can be expressed as an integral along the line-of-sight,
\begin{equation}
\phi({\bf n})=\int_0^{\chi_*}
d\chi\ \Phi(\chi{\bf n})D(\chi),\label{realPhi}
\end{equation}
and it is integrated from 0 to the comoving distance at the last scattering surface $\chi_*$.
Here $\chi(z)$ is the comoving distance at redshift $z$. The galaxy overdensity is also given by a line-of-sight integration as
\begin{equation}
g({\bf n})=\frac{\int d\chi\ b_g\, \mathcal {N}(\chi)\delta(\chi{\bf
n})}{\int d\chi\ \mathcal {N}(\chi)}.\label{realG}
\end{equation}

To better understand the projected galaxy overdensity, $\mathcal {N}(\chi)=dN/d\chi$ is the comoving distance distribution of the galaxies. For NVSS galaxies,  there is a lack of accurate photometric redshifts so approximations to the redshift distributions are made~\cite{Xia:2011hj,Ho:2008bz,Schiavon:2012fc}. Following ~\cite{Smith:2007rg}, we use a Gaussian distribution
\begin{equation}
\frac{dN}{dz} \propto e^{-\frac{(z-z_0)^2}{2\sigma^2}}
\label{redshiftdistri}
\end{equation}
where $\sigma=0.8$ for $z<z_0 (=1.1)$, and $\sigma=0.3$ for $z>z_0$, and the comoving distribution $dN/d\chi$ is easily derived from Eq. (\ref{redshiftdistri}).

The parameter $b_g(z)$ is the redshift-dependent galaxy bias. To keep the model as simple as possible we treat the galaxy bias as a constant which can be determined from a fit to the data shown in Fig.~\ref{nvsscl}. This is different from~\cite{Hirata:2008cb,Ho:2008bz} which used the cross correlation of NVSS with the SDSS and with sources from the 2-Micron All Sky Survey to determine the galaxy bias. The galaxy bias is of great importance because the cross correlation can be directly translated into primordial non-Gaussianity~\cite{NbodyfNL} and may enable general relativity to be tested on cosmological scales~\cite{GR}.

Equations (\ref{realPhi}) and (\ref{realG}) give the general definitions for the lensing potential and galaxy overdensity which are both Gaussian fields, characterized by their variances, (i.e.\ the power spectra) $C_l^{\phi\phi}$ and $C_l^{gg}$. From Eq. (\ref{realPhi}) and (\ref{realG}), one sees that the lensing-galaxy cross correlation is built on the relation described by Eq. (\ref{possion}). Using the Limber approximation $k\sim l/\chi$ we calculate the theoretical galaxy auto-power spectrum and cross-power spectrum in Eqs.~(\ref{clgg}) and (\ref{clphig}).

The galaxy-galaxy power spectrum ( $\langle g_{lm}g^{\ast}_{lm}\rangle$ ) is
\begin{eqnarray}
C_l^{gg}&\simeq&\left(\frac{1}{\int d\chi \mathcal {N}(\chi)}\right)^2 \int d\chi\ b_g^2\, 
\frac{\mathcal
{N}^2(\chi)}{\chi^2}P_{\delta}\left(\frac{l}{\chi}\right)\label{clgg}
\end{eqnarray}
and it will be used later for determining the galaxy bias and for simulating galaxy maps. This power spectrum shows the galaxy clustering strength on different angular scales.  We calculate the power spectrum of the NVSS overdensity map using two independent methods: a pseudo-$C_l$ method and a spherical harmonic estimation, as described in Refs.~\cite{Xia:2011hj, Blake:2004dg} respectively. We find {that both methods agree very well (Fig.~\ref{nvsscl}). As a final check, we have computed the NVSS galaxy power spectrum using the NVSS galaxy map in both equatorial and galactic coordinates and find a negligible difference, as expected, since the galaxy clustering is an intrinsic property of the Universe and does not depend on the choice of coordinate system. The galaxy power spectrum we use, obtained using the pixelized map in equatorial coordinates, is plotted in Fig.~\ref{nvsscl}.

The lensing-galaxy cross-power spectrum is
\begin{eqnarray}
C_l^{\phi g}&\simeq&\frac{3H_0^2\Omega_m}{2}\frac{1}{\int d\chi\
\mathcal {N}(\chi)}\nonumber\\&\times&\int d\chi\
b_g(1+z)D(\chi) \frac{\mathcal
{N}(\chi)}{k^2 \chi^2}P_{\delta}\left(\frac{l}{\chi}\right)
\label{clphig}
\end{eqnarray}
which shows the mutual influence between the gravitational potential and the galaxy clustering in the late universe on different angular scales. $P_{\delta}(k)$ is the matter power spectrum defined using the same convention as Ref.~\cite{pk}.  The primordial scalar curvature perturbations are evaluated at the pivot scale $k_0 = 0.002\, \textrm{Mpc}^{-1}$. The cross-power spectrum $C_l^{\phi g}$ will be used to simulate the correlated galaxy maps and will also be fit to data to determine the detection significance.

\begin{figure}
\includegraphics[bb=50 195 465 480,width=8.5cm]{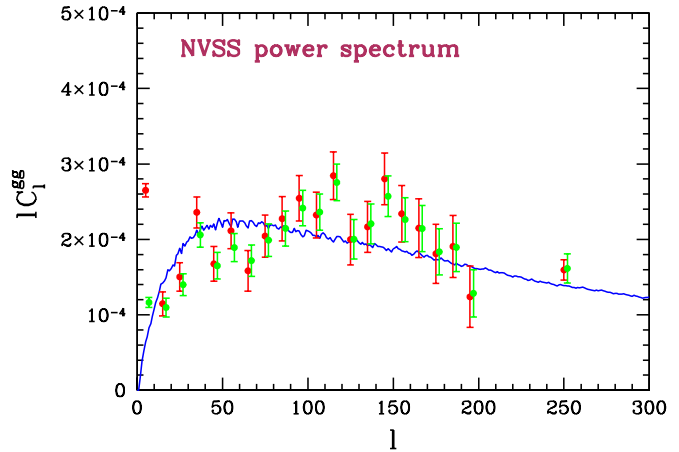}
\caption{The NVSS galaxy auto-power spectrum. The $1\sigma$ error bars are from 1000 Monte Carlo simulations of the NVSS galaxy map with galaxy bias $b_g=1$. For both sets of data points, the red points are from the pseudo-$C_l$ method~\cite{Xia:2011hj}, and the green are from the spherical harmonic estimation~\cite{Blake:2004dg}. The theoretical galaxy auto-power spectrum is fit to the red data points derived from the pseudo-$C_l$ method. Both methods show a consistent galaxy auto-power spectrum from the NVSS data. The first bin of the red data points largely deviates from the theoretical curve due to systematic effects.}
\label{nvsscl}
\end{figure}

\section{Cross Correlation Estimation}
\label{sec:cross}

\begin{table*}
\caption{The 6-parameter $\Lambda$CDM model used for the simulations of the temperature, galaxy and lensing potential. The derived parameter $\sigma_8$, based on the 6-parameter model, is shown in column eight. Using these parameters 1000 galaxy simulations with $b_g=1$ were performed to get the reconstructed galaxy biases as well as the $1\sigma$ error bars. From column nine, we see that all the reconstructed galaxy biases are consistent with the input value $b_g=1$. Furthermore, the galaxy biases of the real data are calculated based on the simulations and shown in column ten. Two independent methods were used to calculate the galaxy auto-power spectra, as specified in the footnotes. }
\begin{tabular}{c|c|c|c|c|c|c|c|c|c}
\hline Data
set&$\Omega_bh^2$&$\Omega_{\rm CDM}h^2$&$H_0$&$A_s$&$n_s$&$\tau$&$\sigma_8$&$b^{\rm sim}_g$&$b^{\rm data}_g$\\
\hline
\multirow{2}{*}{}
WMAP-1\footnotemark[3]&0.0226&0.1104&72&$2.212\times10^{-9}$&0.96&0.117&0.76&$0.98\pm0.12$\footnotemark[1]&$2.00\pm0.12$\footnotemark[1]\\
&&&&&&&&$0.96\pm0.10$\footnotemark[2]&$1.95\pm0.10$\footnotemark[2]\\
\hline
\multirow{2}{*}{}WMAP-3\footnotemark[4]&0.02186&0.1105&70.4&$2.393\times10^{-9}$&0.947&0.073&0.77&$0.98\pm0.11$\footnotemark[1]&$1.97\pm0.11$\footnotemark[1]\\
&&&&&&&&$0.96\pm0.09$\footnotemark[2]&$1.92\pm0.09$\footnotemark[2]\\
\hline
\multirow{2}{*}{}WMAP-5\footnotemark[5]&0.02305&0.1182&69.7&$2.484\times10^{-9}$&0.969&0.094&0.85&$0.98\pm0.10$\footnotemark[1]&$1.85\pm0.10$\footnotemark[1]\\
&&&&&&&&$0.96\pm0.08$\footnotemark[2]&$1.81\pm0.08$\footnotemark[2]\\
\hline
\multirow{2}{*}{}WMAP-7\footnotemark[6]&0.02258&0.1109&71&$2.43\times10^{-9}$&0.963&0.088&0.80&$0.98\pm0.11$\footnotemark[1]&$1.91\pm0.11$\footnotemark[1]\\
&&&&&&&&$0.96\pm0.09$\footnotemark[2]&$1.86\pm0.09$\footnotemark[2]\\
\hline
\end{tabular}
\footnotetext[1]{Pseudo-$C_l$ method~\cite{Xia:2011hj}, }
\footnotetext[2]{Spherical harmonic estimation~\cite{Blake:2004dg}}
\footnotetext[3]{WMAP1+CBI+ACBAR+2dFGRS+Ly$\alpha$, ~\cite{wmap1para}}
\footnotetext[4]{WMAP3 ALL, \url{http://lambda.gsfc.nasa.gov/product/map/dr2/parameters.cfm}}
\footnotetext[5]{WMAP5+BAO+SNALL+Ly$\alpha$POST, \url{http://lambda.gsfc.nasa.gov/product/map/dr3/parameters.cfm}}
\footnotetext[6]{WMAP7, \url{http://lambda.gsfc.nasa.gov/product/map/dr4/parameters.cfm}} \label{paratable}
\end{table*}

Monte Carlo simulations are used to estimate the cross correlation between the CMB and the galaxy distribution. The variances ${\tilde C}_l^{TT}$, $C_l^{TT}$, $C_l^{\phi\phi}$ are computed using \camb~\cite{Lewis:1999bs} with the cosmological parameters listed in Table~\ref{paratable}. In addition to these, we derive $C_l^{gg}$ and $C_l^{\phi g}$ from Eq. (\ref{clgg}) and Eq. (\ref{clphig}) with the parameters listed in Table~\ref{paratable}. Simulated CMB temperature modes, ${\tilde a}_{lm}$, are drawn from Gaussian distributions with zero means and variances ${\tilde C}_l^{TT}$. In this work, we will use two sets of cosmological parameters because we want to check the consistency of our results with a previous study~\cite{Smith:2007rg} and also because we want to explore the impact of using the newest parameters from WMAP-7 on the lensing-galaxy cross correlations. 

We convert these ${\tilde a}_{lm}$ to an unlensed temperature map, ${\tilde T}({\bf n})$, on which we do a cubic interpolation to precisely implement Eq. (\ref{cmblensing}). This produces a lensed temperature map $T({\bf n})$ that is converted back to harmonic space  to give the lensed modes $a_{lm}$. Then each DA's beam and pixel window transfer function (the pixel window transfer function has negligible effects on the cross-power spectra) from WMAP are multiplied by these modes which are subsequently transformed into a temperature map containing the lensing signal.

We then simulate Gaussian noise in map space where the pixel noise is assumed to be uncorrelated and Gaussian distributed with zero mean and pixel-independent variance determined from ${\sigma_0}/{\sqrt{N_{\rm obs}}}$.  Here, both $\sigma_0$ and $N_{\rm obs}$ are supplied by the WMAP team for different DAs. We add this noise map to the signal map and apply the Kp0 mask to get a simulated WMAP DA made in the same way as the real WMAP maps were produced. The entire procedure can be summarized by Eq.~(\ref{simWMAP}) in which $a_{lm}$ is the lensed CMB mode,  $n({\bf n})$ the white noise, ${\rm M}^{\rm WMAP}({\bf n})$ the mask map, $\nu$ the index of the DA channel, $p_l$ the pixel window transfer function, $b_l$ the beam transfer function
\begin{eqnarray}
T^{(\nu)}({\bf n})&=&{\rm M}^{\rm WMAP}({\bf n})\biggl[\sum_{lm}p_lb^{(\nu)}_la_{lm}Y_{lm}({\bf n})\nonumber\\
&&\qquad + \left(\frac{\sigma_0}{\sqrt{N_{\rm obs}({\bf
n})}}\right)^{(\nu)}n({\bf n})\biggr]\,.
\label{simWMAP}
\end{eqnarray}

To maximize the signal-to-noise ratio, we compute a single harmonic mode ${\hat a}_{lm}$ from eight Q, V, W-band DAs. This reduction step is expressed as~\cite{Smith:2007rg}
\begin{eqnarray}
\mathbf{\hat a}&=&(\mathbf{S}+\mathbf{N})^{-1}\mathbf{a}\nonumber\\
&=&\mathbf{S}^{-1/2}\mathbf{A}^{-1}\mathbf{S}^{1/2}\mathbf{N}^{-1}\mathbf{a}.\label{DAto1}
\end{eqnarray}
Here ${\bf a}$ is the vector of $a_{lm}$s, ${\bf S}$ the signal covariance matrix, ${\bf N}$ the noise covariance matrix, and ${\bf A}=\mathbf{I}+\mathbf{S}^{1/2}\mathbf{N}^{-1}\mathbf{S}^{1/2}$.  We use the second form of Eq.~(\ref{DAto1})  and filter the raw CMB modes using a multigrid-preconditioned-complex conjugate gradient method. The master equation that has to be solved is
\begin{equation}
\mathbf{A}\mathbf{x}=\mathbf{y},
\label{eq13}
\end{equation}
where $\mathbf{x}=\mathbf{S}^{1/2}\mathbf{\hat a}$, and
$\mathbf{y}=\mathbf{S}^{1/2}\mathbf{N}^{-1}\mathbf{a}$.
Equation~(\ref{eq13}) is better for numerical computations because ${\bf A}$ is close to the unit matrix. Appendix~\ref{app:mpcg} gives details of the numerical calculation of Eq.~(\ref{eq13}). We solve Eq.~(\ref{eq13}) with $\hat a_{lm}$ for both the temperature ($\hat T_{lm}$) and the galaxy ($\hat g_{lm}$).

We use the standard quadratic estimator to reconstruct a noisy lensing potential map $\hat\phi_{lm}$ in harmonic space~\cite{Hu:2001fa,Hu:2001tn},
\begin{equation}
\displaystyle\sum_{lm}\hat\phi_{lm}Y_{lm}({\bf n})=\nabla^i({\
}_0A^T({\bf n})\nabla_i{\ }_0B^T({\bf n})),\label{noisypotential}
\end{equation}
where \begin{equation} {}_0A^T(\mathbf{n}) =
\displaystyle\sum_{lm}\hat T_{lm} Y_{lm}(\mathbf{n})\end{equation} and
\begin{equation}
_0B^T({\bf{n}}) = \displaystyle \sum_{lm} \tilde{C}^{TT}_l\hat
T_{lm}  Y_{lm}({\bf{n}}).
\end{equation}
In the above equations, $\nabla_i$ is the gradient operator on a sphere and $\nabla^i=g^{ij}\nabla_j$ . Here $g_{ij}$ is the metric of a sphere.

We also use Monte Carlo simulations for the NVSS galaxy maps. The simulated galaxy modes $g_{lm}$ are drawn from a Gaussian distribution and transformed into a galaxy overdensity map $g({\bf n})$ at \healpix\ resolution $N_{\rm side}=1024$. The galaxy modes must satisfy the correct galaxy-galaxy auto-correlation and lensing-galaxy cross correlation. From these two constraints the simulated galaxy mode must be
\begin{equation}
g_{lm}=\frac{C_l^{\phi
g}}{C_l^{\phi\phi}}\phi_{lm}+\sqrt{C_l^{gg}-\frac{(C_l^{\phi
g})^2}{C_l^{\phi\phi}}}G_{lm},
\end{equation}
where $G_{lm}$ is a complex Gaussian random variable, and $\phi_{lm}$ is inherited from the deflection field in Eq. (\ref{cmblensing}). From this equation we see that the lensing-galaxy correlation is encoded in the first term. A NVSS map is generated where the galaxy number count in each pixel is drawn from a Poisson distribution with mean
\begin{equation}
\lambda({\bf n})=\bar n(1+g({\bf n})).
\end{equation}
The galaxy count map $\lambda({\bf n})$ is used to generate a simulated galaxy overdensity
map $g^{(\text{sim})} ({\bf n})$,
\begin{equation}
g^{(\text{sim})} ({\bf n}) ={\rm{M}}^{\rm{NVSS}}({\bf n})\left[\frac{\lambda({\bf
n})}{\bar n}-1\right],
\end{equation}
where ${\rm{M}}^{\rm{NVSS}}$ is the NVSS mask shown in Fig.~\ref{mask}. $g^{(\text{sim})} ({\bf n})$ automatically contains the shot-noise with the variance $N_l^{gg}={1}/{\bar n}$ for the galaxy overdensity map. We degrade this map to resolution $N_{\rm side}=256$ i.e.\ the same as the real NVSS data. The harmonic mode $g_{lm}^{(\text{sim})}$, which contains the shot-noise, is obtained from $g^{(\text{sim})} ({\bf n})$ and is further filtered using the same procedure as in Eq.~(\ref{DAto1}),
\begin{equation}
\hat g_{lm} =({\bf S}+{\bf N})^{-1}g_{lm}^{(\text{sim})} .\label{rawgmap}
\end{equation}
Here ${\bf S}$ represents the primordial galaxy covariance and ${\bf N}$ is the shot-noise covariance.

We show the noisy reconstruction of the potential maps and the filtered galaxy map in Fig.~\ref{tracermap}, using the measured WMAP and NVSS data.
\begin{figure*}
\includegraphics[width=8cm]{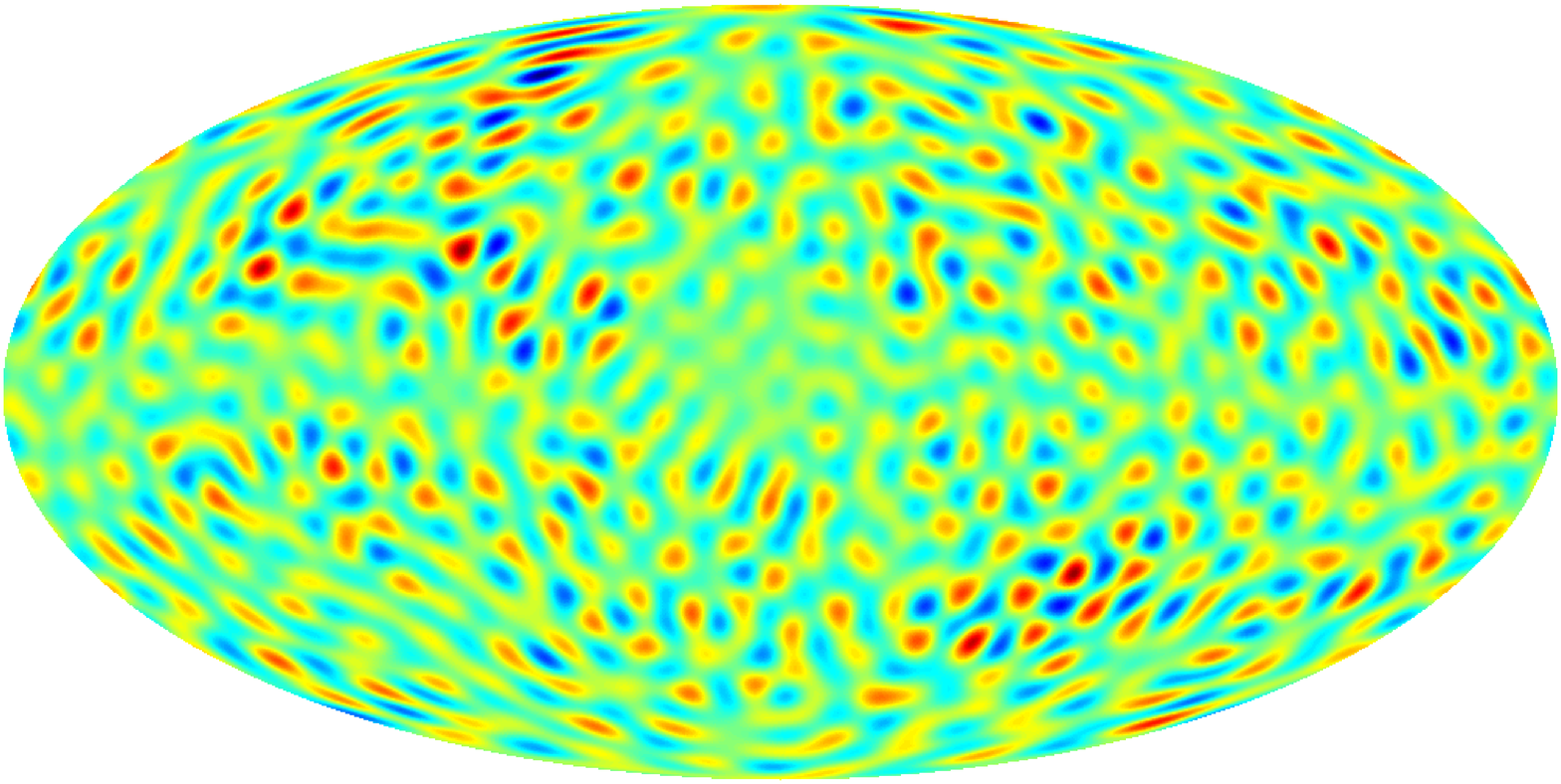}
\vspace{0.5cm}
\includegraphics[width=8cm]{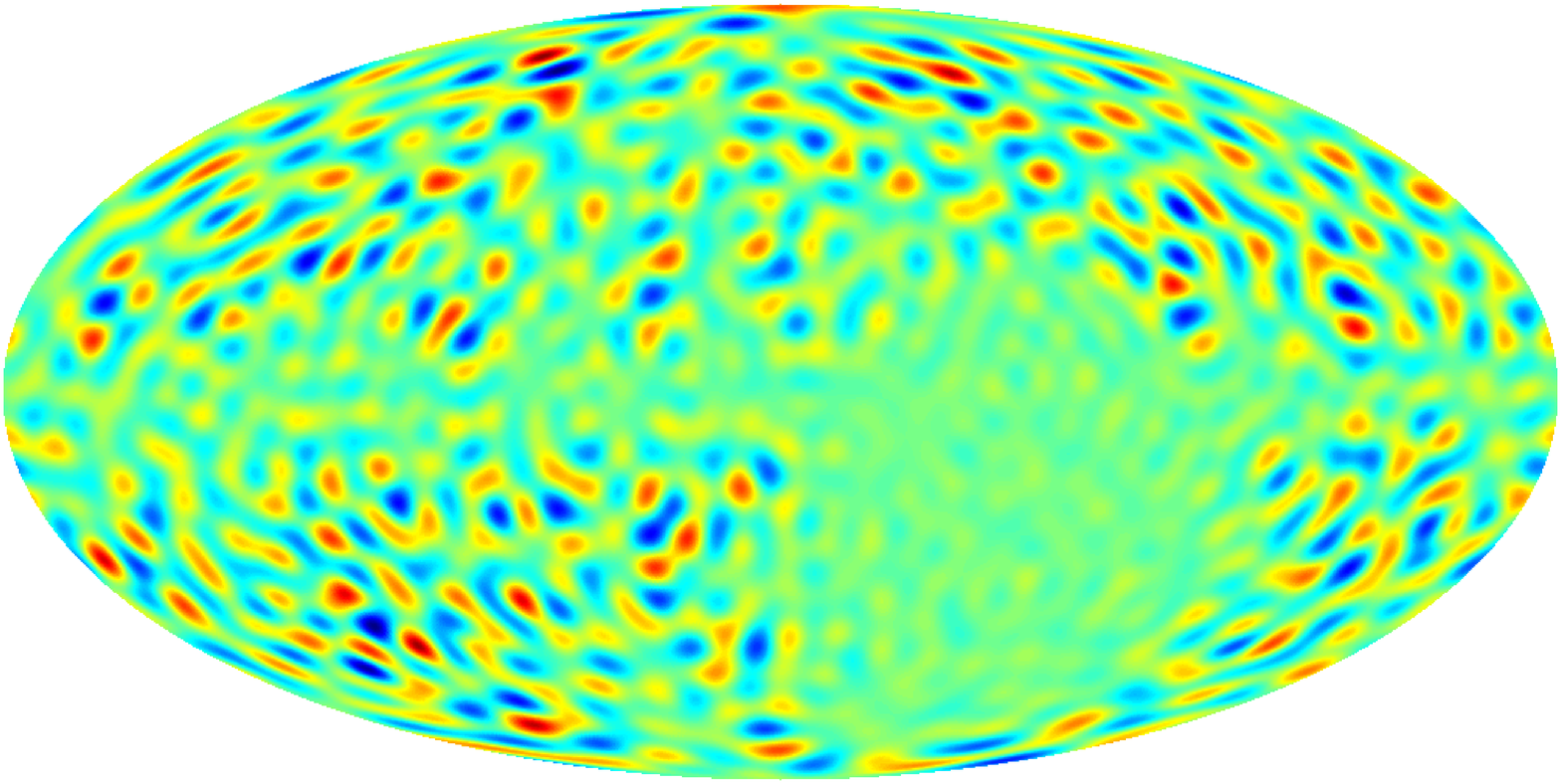}
\caption{The noisy reconstruction of the lensing potential map
(Eq.(\ref{noisypotential}) from WMAP-7) band-pass filtered from $20\le l\le 40$ (left). The analogous map from NVSS galaxy data [Eq.(\ref{rawgmap})] within the band $20\le l\le 40$ (right). } \label{tracermap}
\end{figure*}

The lensing-galaxy cross-power spectrum is the observable which will be compared with the counterpart from data. The estimator of the lensing-galaxy cross correlation is expressed as
\begin{equation}
C^{\phi g}_b=\frac{1}{\mathcal
{F}_b}\displaystyle\sum_{\substack{l\in b\\ -l\le m\le l}} (\hat
\phi_{lm}-\langle\hat\phi_{lm}\rangle)^{\ast}\hat
{g}_{lm},\label{cross-estimator}
\end{equation}
where $\mathcal {F}_b$ is the normalization factor at band $b$. It is shown in Ref.~\cite{Smith:2006ud} that the normalization factor can be calculated by either a 
direct or a fast method for the full-sky coverage and that the fast method converges in $O(10^2)$ simulations. When there is a sky-cut these methods account for the sky-cut effect very well and a constant $f_{\rm sky}$ is often used. The factor $f_{\rm sky}$ is actually a function of $l$~\cite{Verde:2003ey} so a simple constant approximation may potentially bias the cross-spectra reconstruction. Therefore an end-to-end simulation~\cite{Smith:2007rg} is the best way to get the exact normalization accounting for the sky-cut and that is done here.

\begin{table*}
\caption{The two sets of cosmological parameters used in this work:  we choose two sets of parameters (labeled ``\pthree" and ``\pseven") to do the cross correlation calculations in this work. In order to compare our results with those from the previous studies~\cite{Smith:2007rg}, we use the parameters they used \pthree\ from WMAP-3's cosmological parameters ( row ``WMAP-3" in Table \ref{paratable}) combined with the corresponding galaxy bias of Smith \textit{et al}.~\cite{Smith:2007rg}.  Based on the newest cosmological parameters from WMAP-7 ( row ``WMAP-7" in Table \ref{paratable}) we construct a new parameter set, \pseven\ with the corresponding galaxy bias shown in Table \ref{paratable}. }
\begin{tabular}{c|c|c|c|c|c|c|c|c}
\hline Data
set&$\Omega_bh^2$&$\Omega_{\rm CDM}h^2$&$H_0$&$A_s$&$n_s$&$\tau$&$\sigma_8$&$b_g$\\
\hline
\pthree &0.02186&0.1105&70.4&$2.393\times10^{-9}$&0.947&0.073&0.77&$1.70$\\
\hline
\pseven &0.02258&0.1109&71&$2.43\times10^{-9}$&0.963&0.088&0.80&$1.91$\\
\hline
\end{tabular}
\label{paratable1}
\end{table*}

\begin{table*}
\centering \caption{Measure of lensing-galaxy cross correlation
$\mathcal{C}$ and its significance $\mathcal{C} / \Delta
\mathcal{C}$ using  \pthree. For five columns of this table: the second column shows the simulation results, the third column is the case without gradient stripes removed, the fourth column is the case with gradient stripes removed (this column shows the statistical results of the lensing-galaxy cross correlations). The fifth column is the case by setting the NVSS map in equatorial coordinates as a complementary diagnostic test. }
\begin{tabular}{c|cc|cc|cc|c}
\hline\hline
Data set & $\mathcal {C}^{\rm{sim}}$ & $\mathcal{C} / \Delta \mathcal{C}$& $\mathcal {C}$\footnotemark[1]& $\mathcal{C} / \Delta \mathcal{C}$&$\mathcal {C}$\footnotemark[2] & $\mathcal{C} / \Delta \mathcal{C}$& $\mathcal {C}$\footnotemark[3]\\
\hline
WMAP-1$\times$NVSS& $1.00\pm0.47$ & $2.13\sigma$& $1.25\pm0.47$ & $2.66\sigma$&$1.24\pm0.47$&$2.64\sigma$&0.26\\
\hline
WMAP-3$\times$NVSS& $1.00\pm0.35$ & $2.86\sigma$& $1.20\pm0.35$ & $3.43\sigma$&$1.26\pm0.35$&$3.60\sigma$&0.17\\
\hline
WMAP-5$\times$NVSS& $1.00\pm0.31$ & $3.23\sigma$& $1.24\pm0.31$ & $4.00\sigma$&$1.27\pm0.31$&$4.10\sigma$&0.23\\
\hline
WMAP-7$\times$NVSS& $1.00\pm0.30$ & $3.33\sigma$& $1.14\pm0.30$ & $3.80\sigma$&$1.16\pm0.30$&$3.87\sigma$&0.15\\
\hline
\end{tabular} \\
\footnotetext[1]{Without gradient stripes removed.}
\footnotetext[2]{With gradient stripes removed.}
\footnotetext[3]{Galaxy map in equatorial coordinate.}
\label{lensingtable1}
\end{table*}

\begin{table*}
\centering \caption{Measure of lensing-galaxy cross correlation
$\mathcal{C}$ and its significance $\mathcal{C} / \Delta
\mathcal{C}$ using \pseven. The format of this table is the same as Table \ref{lensingtable1}. }
\begin{tabular}{c|cc|cc|cc|c}
\hline\hline
Data set & $\mathcal {C}^{\rm{sim}}$ & $\mathcal{C} / \Delta \mathcal{C}$& $\mathcal {C}$\footnotemark[1]& $\mathcal{C} / \Delta \mathcal{C}$&$\mathcal {C}$\footnotemark[2] & $\mathcal{C} / \Delta \mathcal{C}$& $\mathcal {C}$\footnotemark[3]\\
\hline
WMAP-1$\times$NVSS& $1.00\pm0.41$ & $2.44\sigma$& $1.01\pm0.41$ & $2.46\sigma$&$1.00\pm0.41$&$2.44\sigma$&0.20\\
\hline
WMAP-3$\times$NVSS& $1.00\pm0.31$ & $3.23\sigma$& $0.96\pm0.31$ & $3.10\sigma$&$1.01\pm0.31$&$3.26\sigma$&0.13\\
\hline
WMAP-5$\times$NVSS& $1.00\pm0.28$ & $3.57\sigma$& $0.98\pm0.28$ & $3.50\sigma$&$1.01\pm0.28$&$3.61\sigma$&0.18\\
\hline
WMAP-7$\times$NVSS& $1.00\pm0.26$ & $3.85\sigma$& $0.92\pm0.26$ & $3.54\sigma$&$0.93\pm0.26$&$3.58\sigma$&0.11\\
\hline
\end{tabular} \\
\footnotetext[1]{Without gradient stripes removed.}
\footnotetext[2]{With gradient stripes removed.}
\footnotetext[3]{Galaxy map in equatorial coordinate.}
\label{lensingtable2}
\end{table*}

\begin{table}
\centering \caption{Fisher matrix analysis for WMAP$\times$NVSS cross
correlation. The $1\sigma$ error bars are determined from Eq. (\ref{fisherErrorbar}). We calculate two sets of the optimal bounds for this work, based on two sets of parameters: \pthree\ (column two); \pseven\ (column three). }
\begin{tabular}{c|cc|cc}
\hline
Data set &$\mathcal {C}^{\textrm{optimal}}$\,\footnotemark[1]&$\mathcal{C} / \Delta \mathcal{C}$&$\mathcal {C}^{\textrm{optimal}}$\,\footnotemark[2]&$\mathcal{C} / \Delta \mathcal{C}$\\
\hline
WMAP-1$\times$NVSS&$1\pm0.46$&$2.17\sigma$&$1\pm0.39$&$2.56\sigma$\\
\hline
WMAP-3$\times$NVSS&$1\pm0.29$&$3.45\sigma$&$1\pm0.25$&$4.00\sigma$\\
\hline
WMAP-5$\times$NVSS&$1\pm0.25$&$4.00\sigma$&$1\pm0.21$&$4.76\sigma$\\
\hline
WMAP-7$\times$NVSS&$1\pm0.22$&$4.55\sigma$&$1\pm0.19$&$5.26\sigma$\\
\hline
\end{tabular}\\
\footnotetext[1]{WMAP-3 year cosmological parameters and $b_g=1.70$.}
\footnotetext[2]{WMAP-7 year cosmological parameters and
$b_g=1.91$.}
\label{opboundWMAPNVSS}
\end{table}

\begin{table*}
\centering \caption{Results of the curl null tests for WMAP$\times$NVSS cross correlation. The curl null tests are performed based on two sets of parameters: \pthree\ (column two); \pseven\ (column three).  }
\begin{tabular}{c|cc|cc}
\hline
Data set &$\mathcal {C}$\footnotemark[1]&$\mathcal{C} / \Delta \mathcal{C}$&$\mathcal {C}$\footnotemark[2]&$\mathcal{C} / \Delta \mathcal{C}$\\
\hline
WMAP-1$\times$NVSS&$-0.11\pm0.47$&$-0.23\sigma$&$-0.03\pm0.41$&$-0.07\sigma$\\
\hline
WMAP-3$\times$NVSS&$0.00\pm0.35$&$0.00\sigma$&$0.04\pm0.31$&$0.13\sigma$\\
\hline
WMAP-5$\times$NVSS&$0.05\pm0.31$&$0.16\sigma$&$0.07\pm0.28$&$0.25\sigma$\\
\hline
WMAP-7$\times$NVSS&$-0.05\pm0.30$&$0.17\sigma$&$-0.03\pm0.26$&$-0.12\sigma$\\
\hline
\end{tabular}\\
\footnotetext[1]{WMAP-3 year cosmological parameters and $b_g=1.70$.}
\footnotetext[2]{WMAP-7 year cosmological parameters and
$b_g=1.91$.}
\label{curltable}
\end{table*}

\begin{table*}
\centering \caption{Gaussianity diagnostics for the probability distribution of $\{\mathcal{C}\}$ which is constructed from 1000 Monte Carlo simulations. The second column is the Kolmogorov-Smirnov test, and the critical value is 0.04 at 5\% confidence level. The Kolmogorov-Smirnov test requires the maximum deviation be $<0.04$ to validate the distribution is Gaussian. The third column is the skewness of $\{\mathcal{C}\}$, and the fourth column is the kurtosis of $\{\mathcal{C}\}$. The upper values in the cells are the results for \pthree, the lower values for \pseven.  For a Gaussian distribution, the skewness should be 0 and the kurtosis should be 3. As can be seen, all the probability distribution functions pass the Kolmogorov-Smirnov test and are consistent with being Gaussian-distributed.  }
\begin{tabular}{c|c|c|c}
\hline
Data set &maximum distance[$<$0.04]&skewness[$\sim$0]&kurtosis[$\sim$3]\\
\hline
\multirow{2}{*}
{WMAP-1$\times$NVSS}&0.02&0.02&2.89\\
&0.02&-0.05&2.77\\
\hline
\multirow{2}{*}{WMAP-3$\times$NVSS}&0.02&-0.14&2.62\\
&0.03&-0.17&2.58\\
\hline
\multirow{2}{*}{WMAP-5$\times$NVSS}&0.03&-0.17&2.53\\
&0.03&-0.23&2.43\\
\hline
\multirow{2}{*}{WMAP-7$\times$NVSS}&0.03&-0.21&2.44\\
&0.03&-0.21&2.36\\
\hline
\end{tabular}\\
\label{kstable}
\end{table*}

As a systematic check we note that the lensing signal consists of a gradient and a curl component~\cite{curl2005}. The curl component estimator $\psi_{lm}$ is
defined by
\begin{equation}
\displaystyle\sum_{lm}\psi_{lm}Y_{lm}({\bf n})=\displaystyle
\sum_{ij} \epsilon^{ij}\nabla_i({\ }_0A^T({\bf n})\nabla_j{\
}_0B^T({\bf n}))\label{curl-estimator}
\end{equation}
and should vanish because lensing does not generate vorticity. Similar to Eq. (\ref{cross-estimator}), the curl-galaxy
cross correlation diagnostic is calculated by
\begin{equation}
C^{\psi g}_b=\frac{1}{\mathcal
{F}_b}\displaystyle\sum_{\substack{l\in b\\ -l\le m\le l}} (\psi_{lm}-\langle\psi_{lm}\rangle)^{\ast}\hat
{g}_{lm}\label{curl-cross}
\end{equation}
which should also vanish.

The amplitude of the cross correlation is determined using
\begin{equation}
\mathcal {C} = \frac{\sum_{AB} C^{(\textrm{th})}_A
\mathbf{C}^{-1}_{AB} C^{(\textrm{obs})}_{B}}
{\sum_{AB}C^{(\textrm{th})}_A\mathbf{C}^{-1}_{AB}C^{(\textrm{th})}_{B}}.\label{CL}
\end{equation}
${\bf C}_{AB}$ is the covariance matrix for the band powers and $A$ and $B$ stand for the band power index. We find that the off-diagonal correlations of ${\bf C}_{AB}$ are negligible, and the covariance matrix elements can be simply replaced by the band power variance $\sigma_A^2$, i.e.\ ${\bf C}_{AB}=\sigma_A^2\delta_{AB}$.

We have described the procedures used to perform analysis on simulated or measured data. Now we summarize the analysis of the real WMAP and NVSS data.
We fit the theoretical galaxy auto-power spectrum to the NVSS data in Fig.~\ref{nvsscl} and determine the galaxy biases (Table \ref{paratable}) using two methods. The error bars are determined from 1000 simulated galaxy maps with galaxy bias $b_g=1$. Then we choose two sets of parameters (labeled ``\pthree" and ``\pseven") in Table~\ref{paratable1} to do the cross correlation calculations in this work.  In order to compare our results with those from the previous studies~\cite{Smith:2007rg}, we use the parameters they used \pthree\ from WMAP-3's cosmological parameters ( row ``WMAP-3" in Table \ref{paratable}) combined with the corresponding galaxy bias of Smith \textit{et al}.~\cite{Smith:2007rg}.  Based on the newest cosmological parameters from WMAP-7 ( row ``WMAP-7" in Table \ref{paratable}) we construct a new parameter set \pseven\ with the corresponding galaxy bias shown in Table \ref{paratable}. For each of the parameter sets we calculate four lensing-galaxy cross correlations from WMAP-1 to WMAP-7.

We carefully treat the known systematics of NVSS, i.e.\ the gradient stripes which are generated by the declination-dependence of the galaxy overdensity field due to the low-flux calibration issue~\cite{Condon:1998iy}. We first make a gradient stripe map only using $m=0$ modes and then subtract it from the galaxy map. We calculate the lensing-galaxy cross correlations for two cases: without the gradient stripes removed and with the gradient stripes removed. We find that this systematic effect does not affect the lensing-galaxy cross correlations as seen from column $\mathcal C^a$ and column $\mathcal C^b$ in Table~\ref{lensingtable1} and Table~\ref{lensingtable2}. The statistical results are those with the gradient stripes removed which are shown in Figs.~\ref{crosspower} and ~\ref{crosspowerpara2} for the Kp0$+$NVSS mask combination. From the two figures, we find that the lensing-galaxy cross-power spectra are consistent with the theoretical predictions and the uncertainty of the cross-power spectrum is decreasing as the year of WMAP increases. All the error bars are calculated from 1000 Monte Carlo simulations, which we confirmed to be sufficient for convergences. As a complementary diagnostic test, we keep the NVSS galaxy overdensity map in equatorial coordinates and calculate the cross-power spectra and all the amplitudes are shown in column $\mathcal C^c$ in Table~\ref{lensingtable1} and Table~\ref{lensingtable2}. As can be seen, they are negligible. All the cross correlation amplitudes are summarized in Table~\ref{lensingtable1} and Table~\ref{lensingtable2}. From the results of  WMAP-3$\times$NVSS in Table~\ref{lensingtable1}: for the statistical results, we get lensing detection significance level of $3.60\sigma$ and ~\cite{Smith:2007rg} got $4.02\sigma$. Both analyses agree quite well. We find the cross-power spectra from WMAP-5 and WMAP-7 clearly and firmly show the lensing-galaxy correlation at $>3\sigma$ level for both cases. All the results are within the optimal bounds shown in Table~\ref{opboundWMAPNVSS}.

Assuming there is no cosmological parity violation the curl-galaxy correlation should be consistent with zero. We show the results of the curl null tests in Figs.~\ref{curlpower} and \ref{curlpowerpara2}. As expected, all the correlations are consistent with zero. The amplitude as well as the significance are given in Table~\ref{curltable}.

We pixelized the NVSS catalog with different \healpix\ resolutions (e.g. $N_{\rm side}=512, 1024$)  to probe the possible pixel artifacts that could afflict the cross-power spectra and because we want to examine the impact of possible long range spatial correlations. However, we find that different NVSS pixelization resolutions do not affect the cross correlation.

We also use the diagonal elements of the covariance matrix to do the analysis to check the consistency with previous studies. In this case, the estimator has a larger variance as pointed out by Smith \textit{et al}.~\cite{Smith:2009jr}. This contributes to the difference in significance levels between $4\sigma$~\cite{Smith:2007rg} and $2\sigma$~\cite{Hirata:2008cb}.

\begin{figure*}
\includegraphics[bb=50 195 465 480,width=8.5cm]{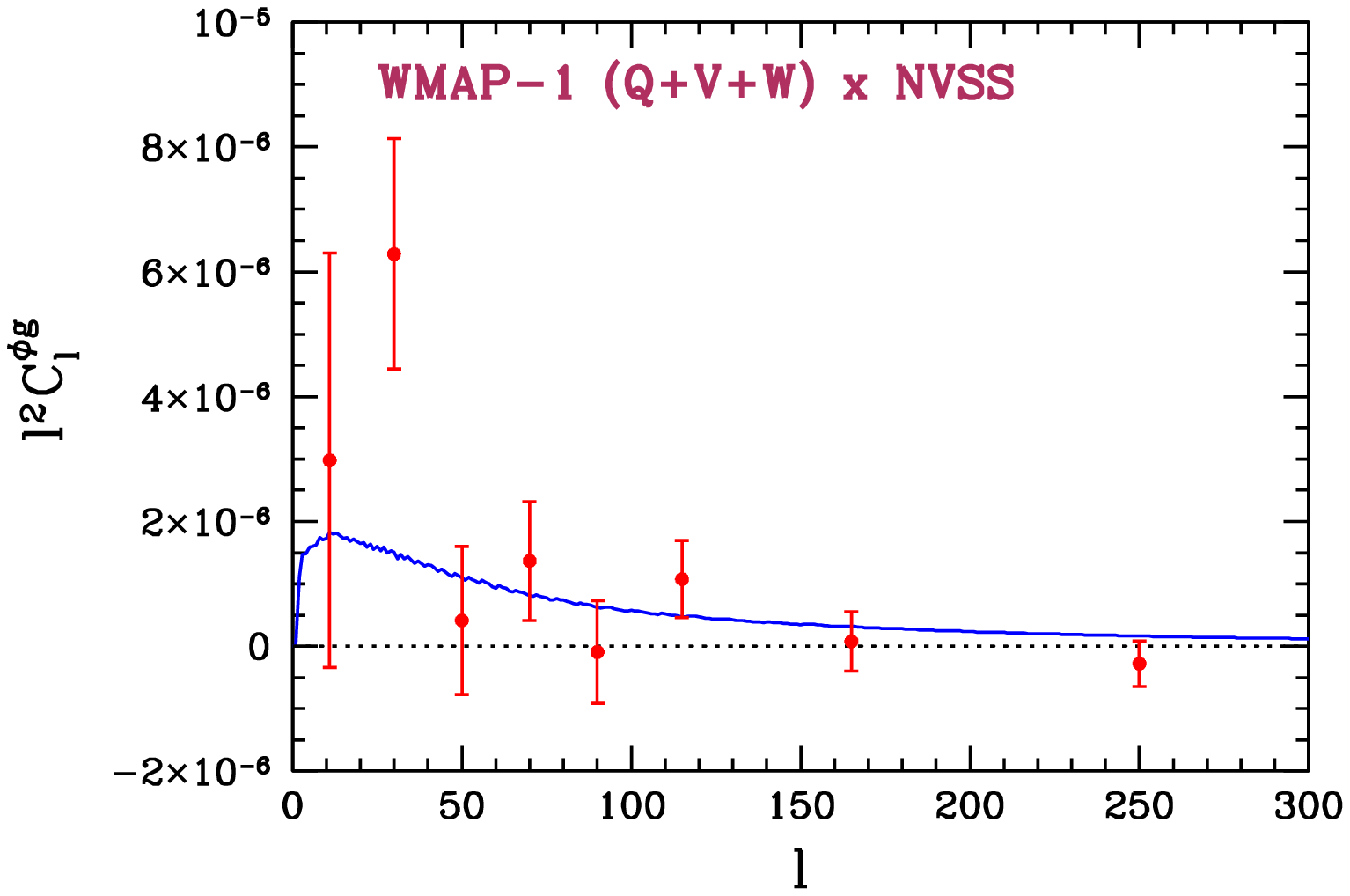}\qquad
\includegraphics[bb=50 195 465 480,width=8.5cm]{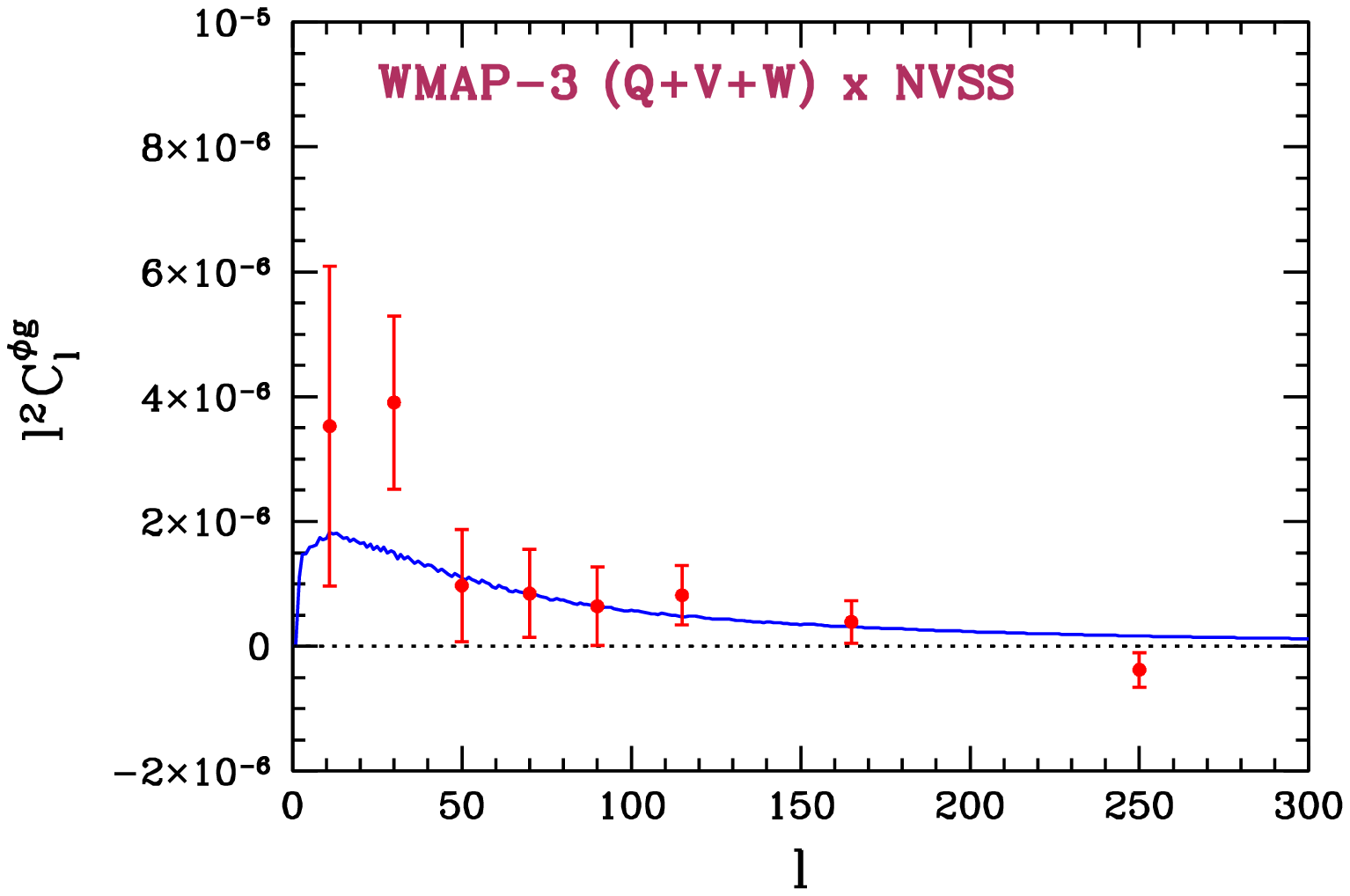} \\
\vspace{0.5cm}
\includegraphics[bb=50 195 465 480,width=8.5cm]{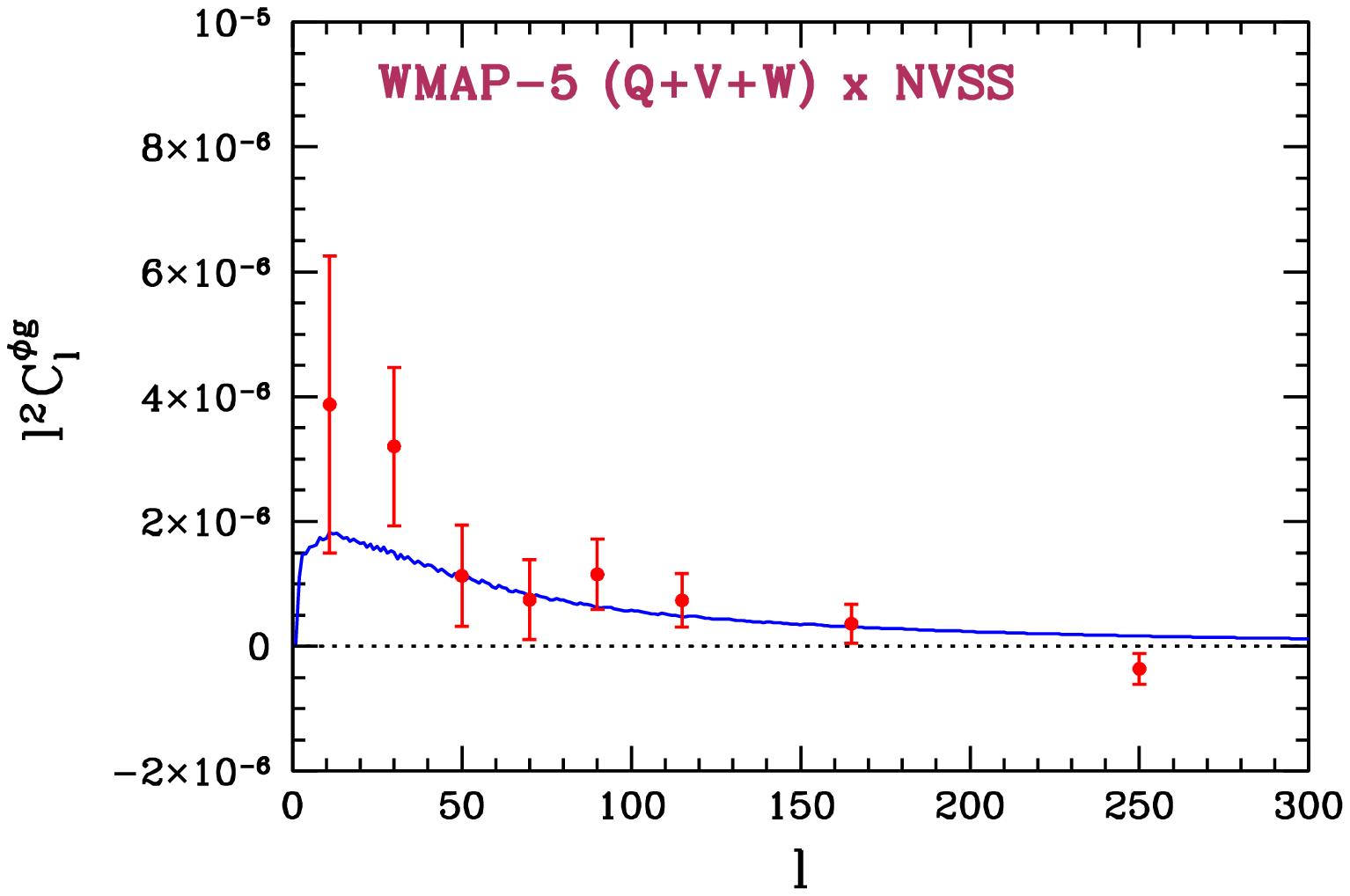}\qquad
\includegraphics[bb=50 195 465 480,width=8.5cm]{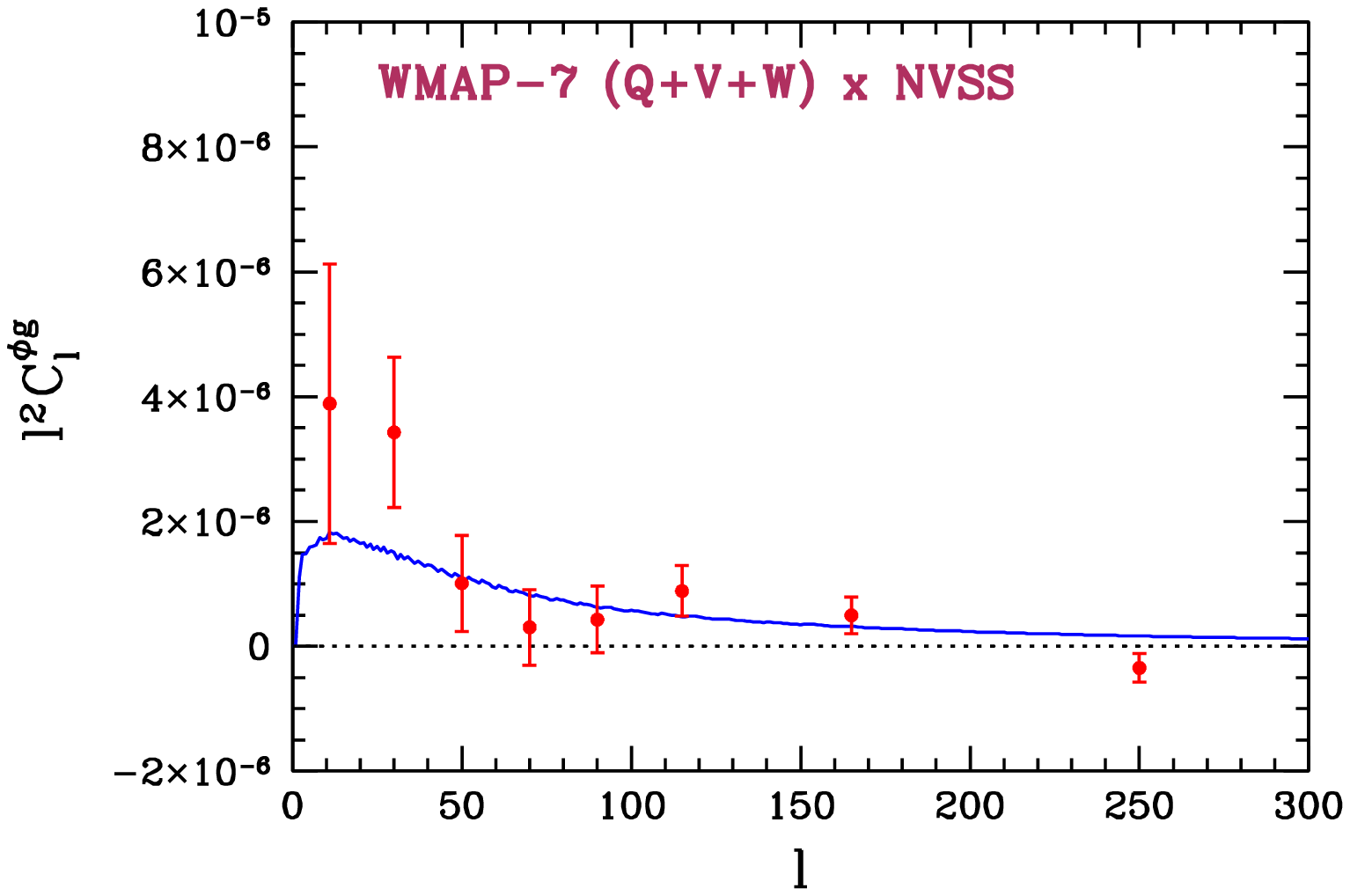}
\caption{(\pthree) The lensing-galaxy cross-power spectra for WMAP$\times$ NVSS are calculated from Eq. (\ref{cross-estimator}). The Kp0 mask is used to remove the contaminated regions of the WMAP data. WMAP's data are provided from two Q bands, two V bands and four W bands. The NVSS mask is applied to the galaxy map to remove bright sources and unobserved regions. The theoretical cross-power spectra are shown in blue solid lines, and they are the same for all of the four panels. The real data are shown in the red scattered points. The statistical amplitude for WMAP-1$\times$NVSS is $1.24\pm0.47$, for WMAP-3$\times$NVSS is $1.26\pm0.35$, for WMAP-5$\times$NVSS is $1.27\pm0.31$, for WMAP-7$\times$NVSS is $1.16\pm0.30$. All the error bars are determined from 1000 Monte Carlo simulations. We find that the lensing-galaxy cross-power spectra are consistent with the theoretical predictions and the uncertainty of the cross-power spectrum is decreasing as the year of WMAP increases. }
\label{crosspower}
\end{figure*}
\begin{figure*}
\includegraphics[bb=50 195 465 480,width=8.5cm]{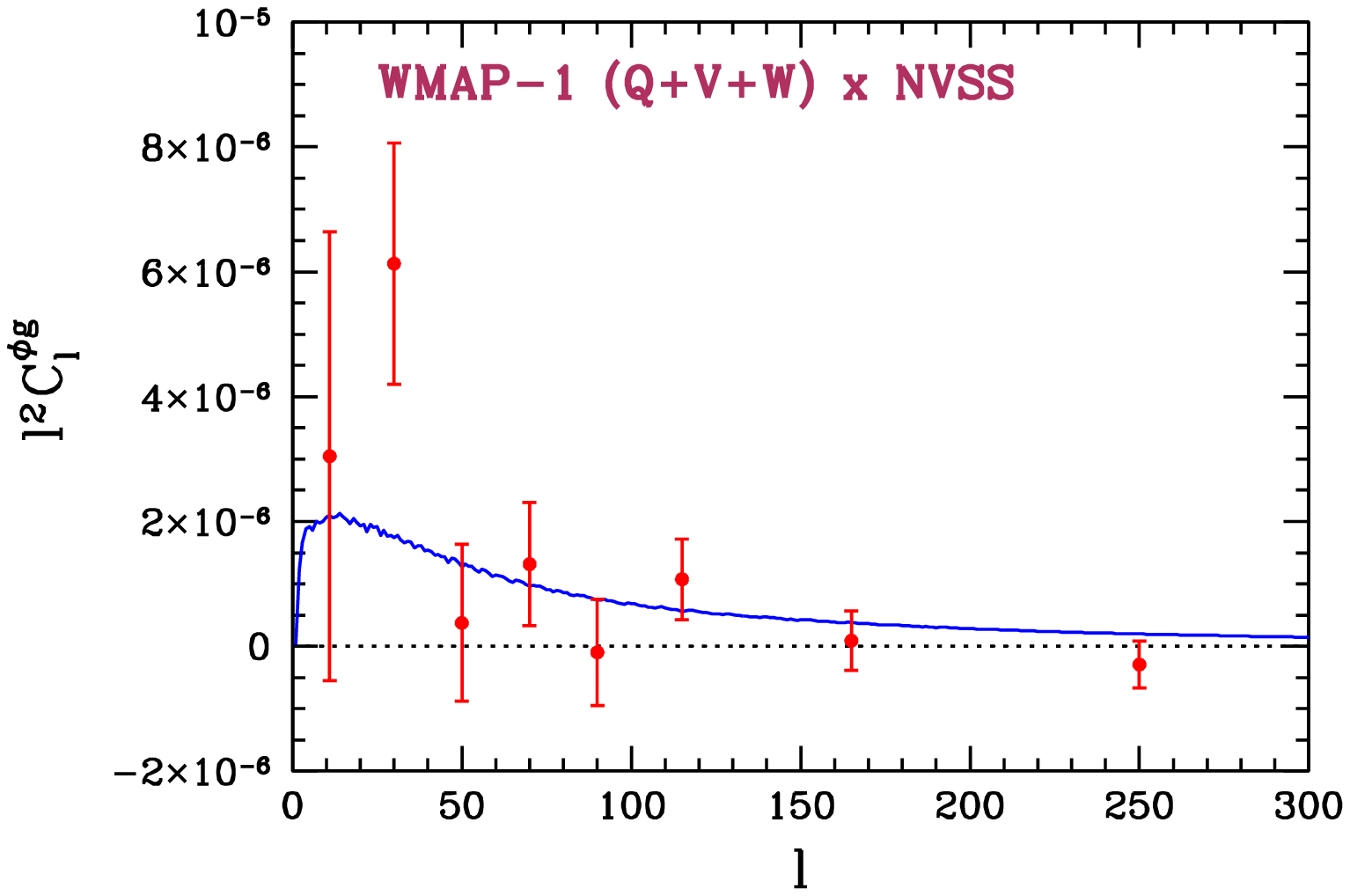}\qquad
\includegraphics[bb=50 195 465 480,width=8.5cm]{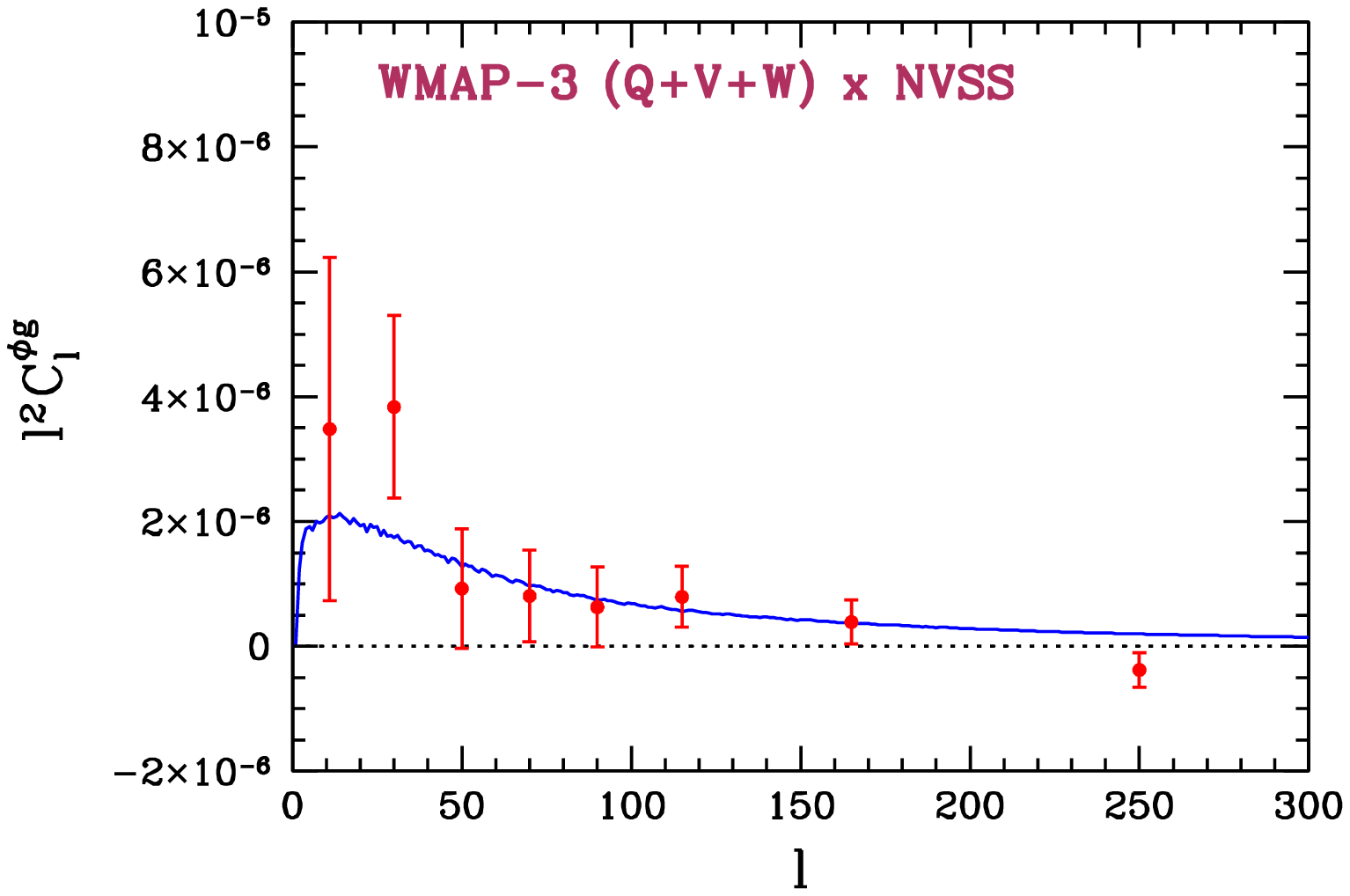} \\
\vspace{0.5cm}
\includegraphics[bb=50 195 465 480,width=8.5cm]{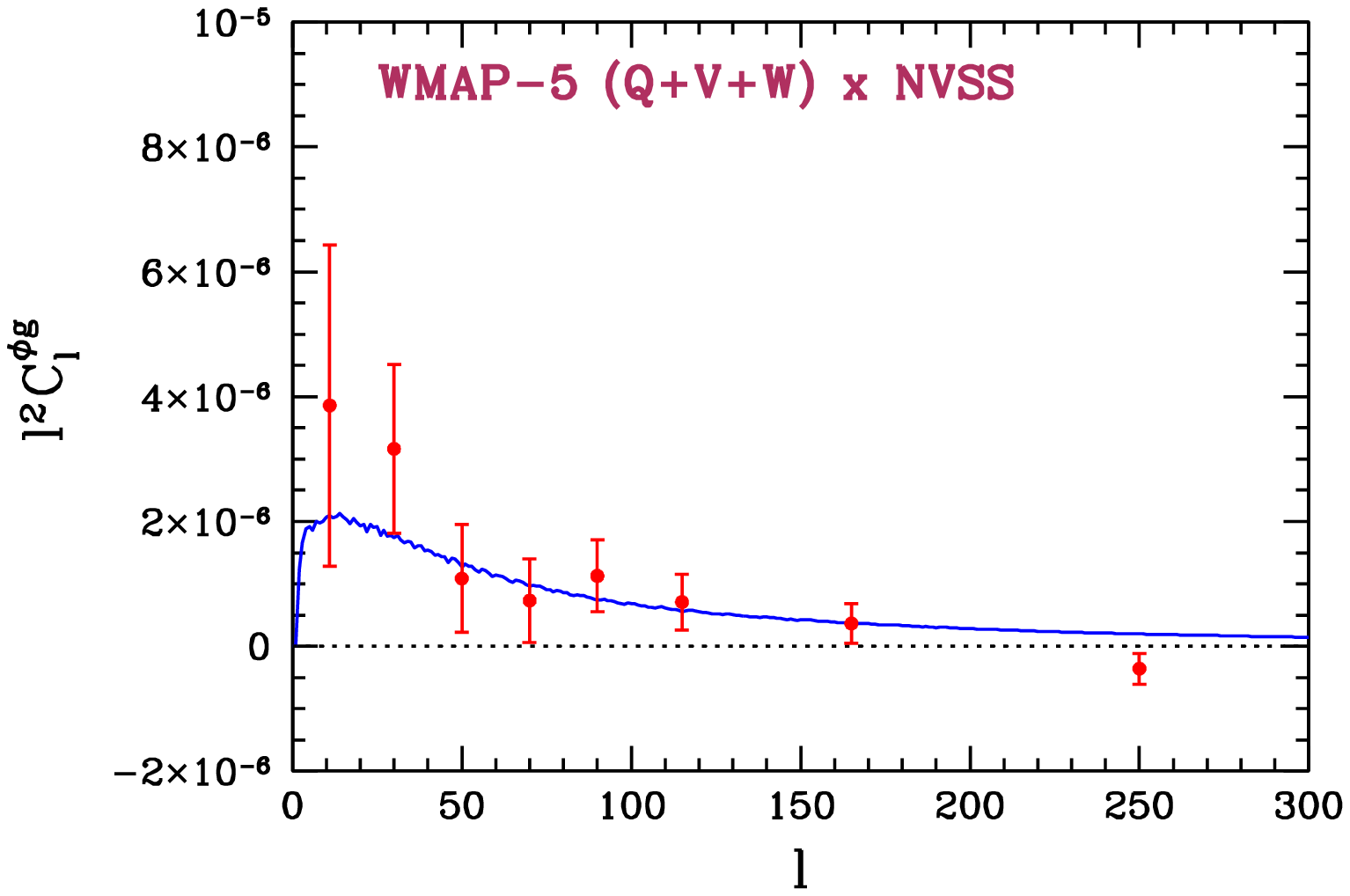}\qquad
\includegraphics[bb=50 195 465 480,width=8.5cm]{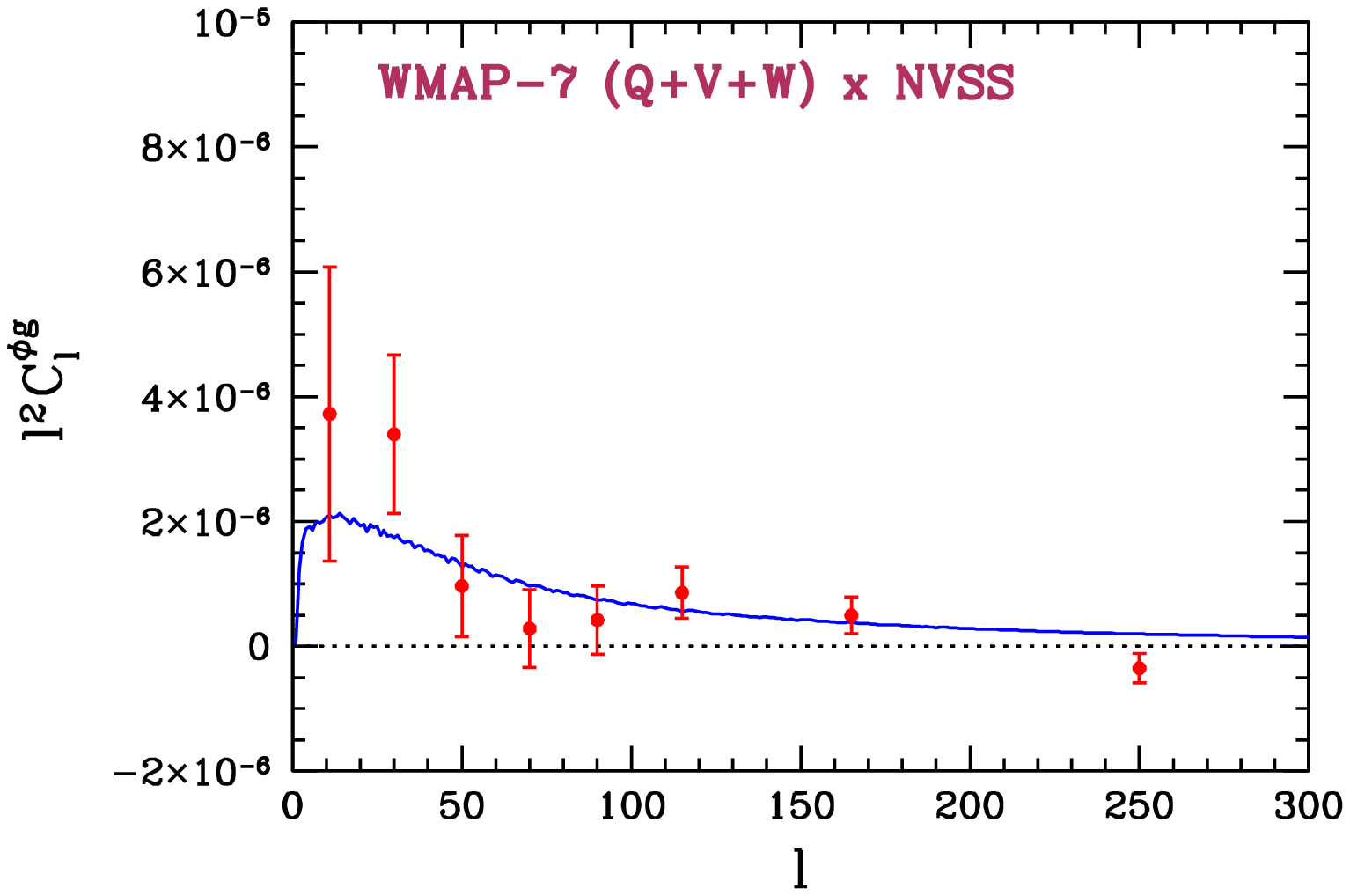}
\caption{(\pseven) The lensing-galaxy cross-power spectra for WMAP$\times$ NVSS are calculated from Eq. (\ref{cross-estimator}). See Fig.~\ref{crosspower} for detailed descriptions. The statistical amplitude for WMAP-1$\times$NVSS is $1.00\pm0.41$, for WMAP-3$\times$NVSS is $1.01\pm0.31$, for WMAP-5$\times$NVSS is $1.01\pm0.28$, for WMAP-7$\times$NVSS is $0.93\pm0.26$. }
\label{crosspowerpara2}
\end{figure*}

\begin{figure*}
\includegraphics[bb=50 195 465 480,width=8.5cm]{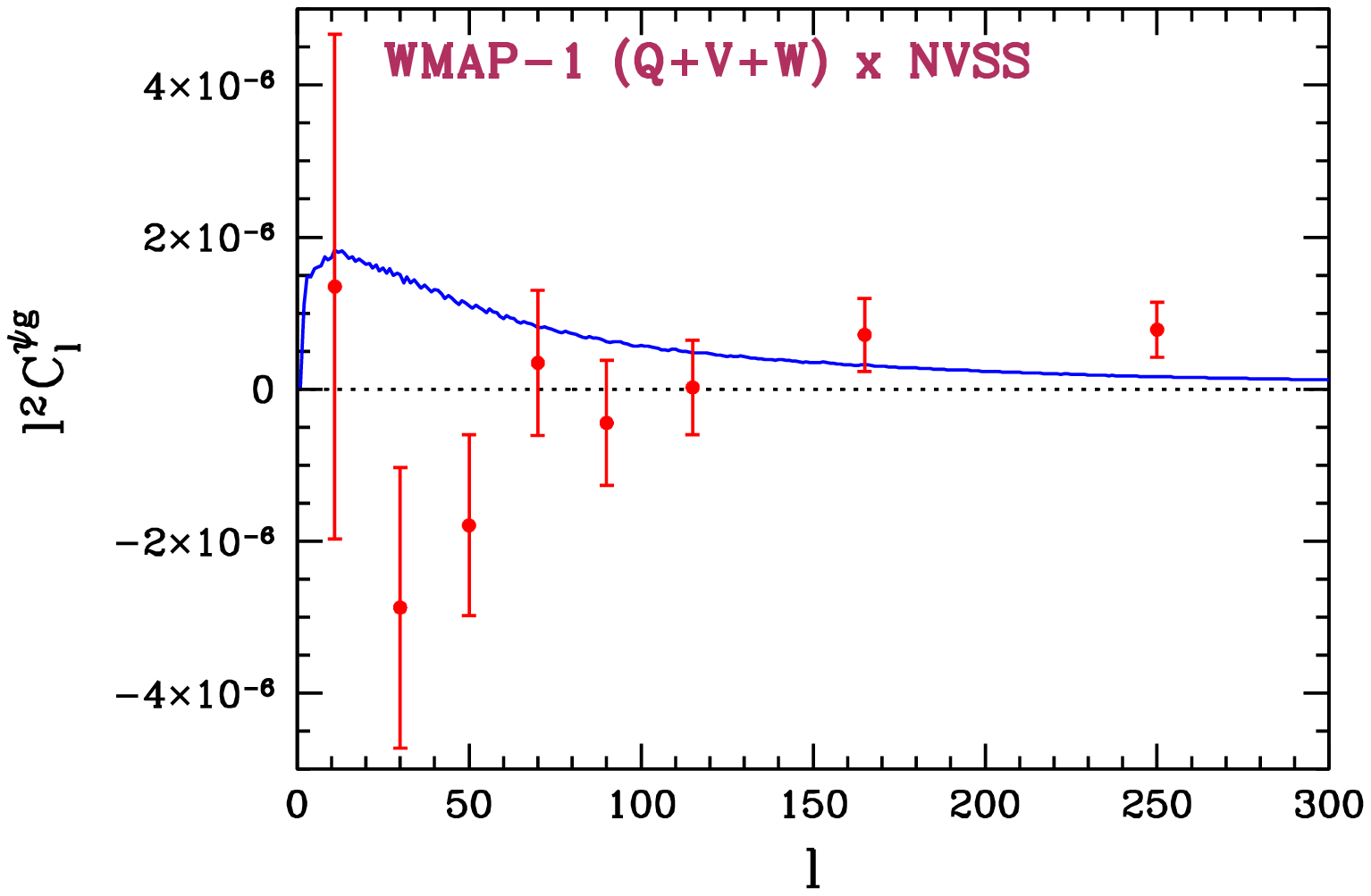}\qquad
\includegraphics[bb=50 195 465 480,width=8.5cm]{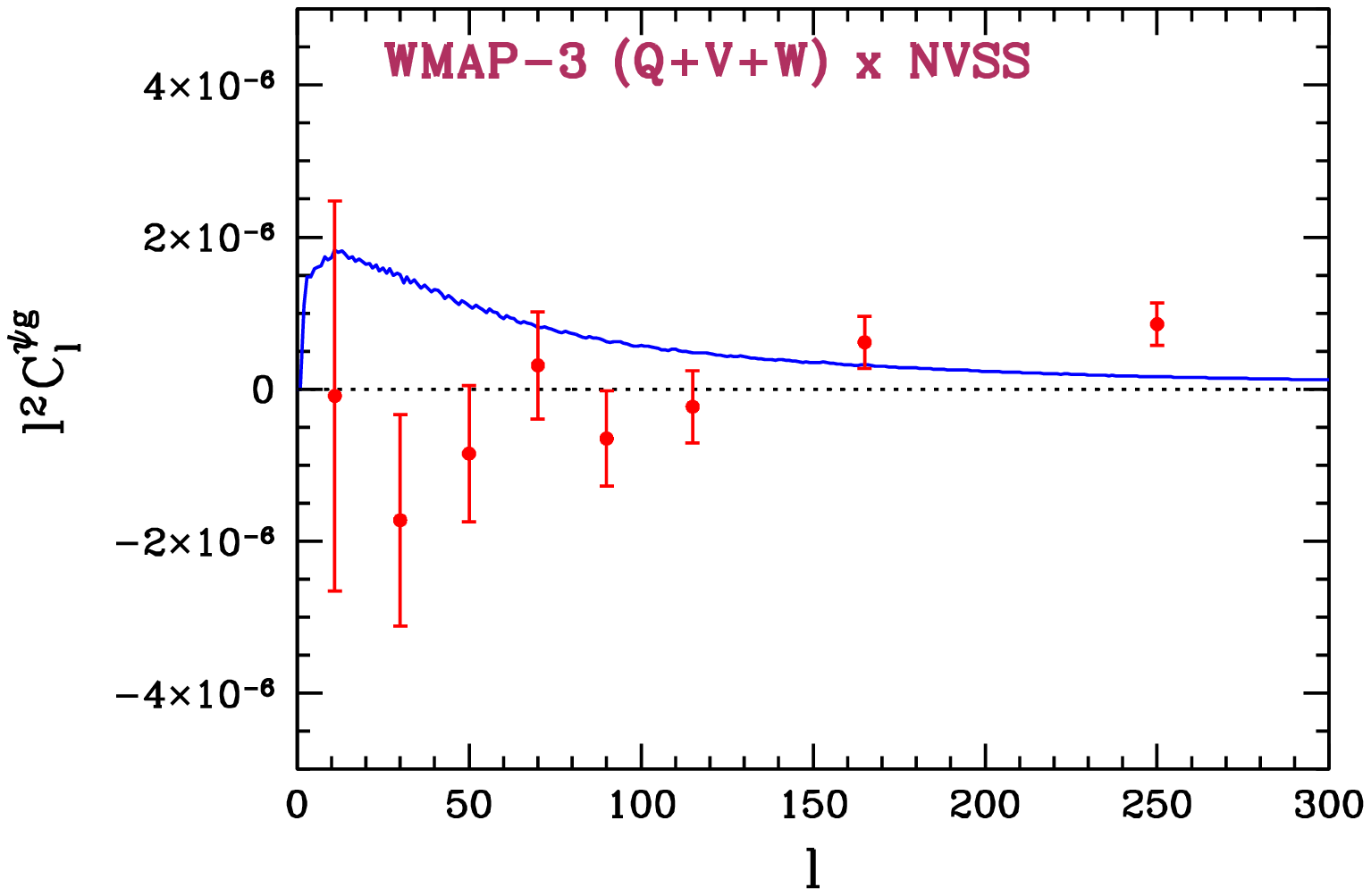} \\
\vspace{0.5cm}
\includegraphics[bb=50 195 465 480,width=8.5cm]{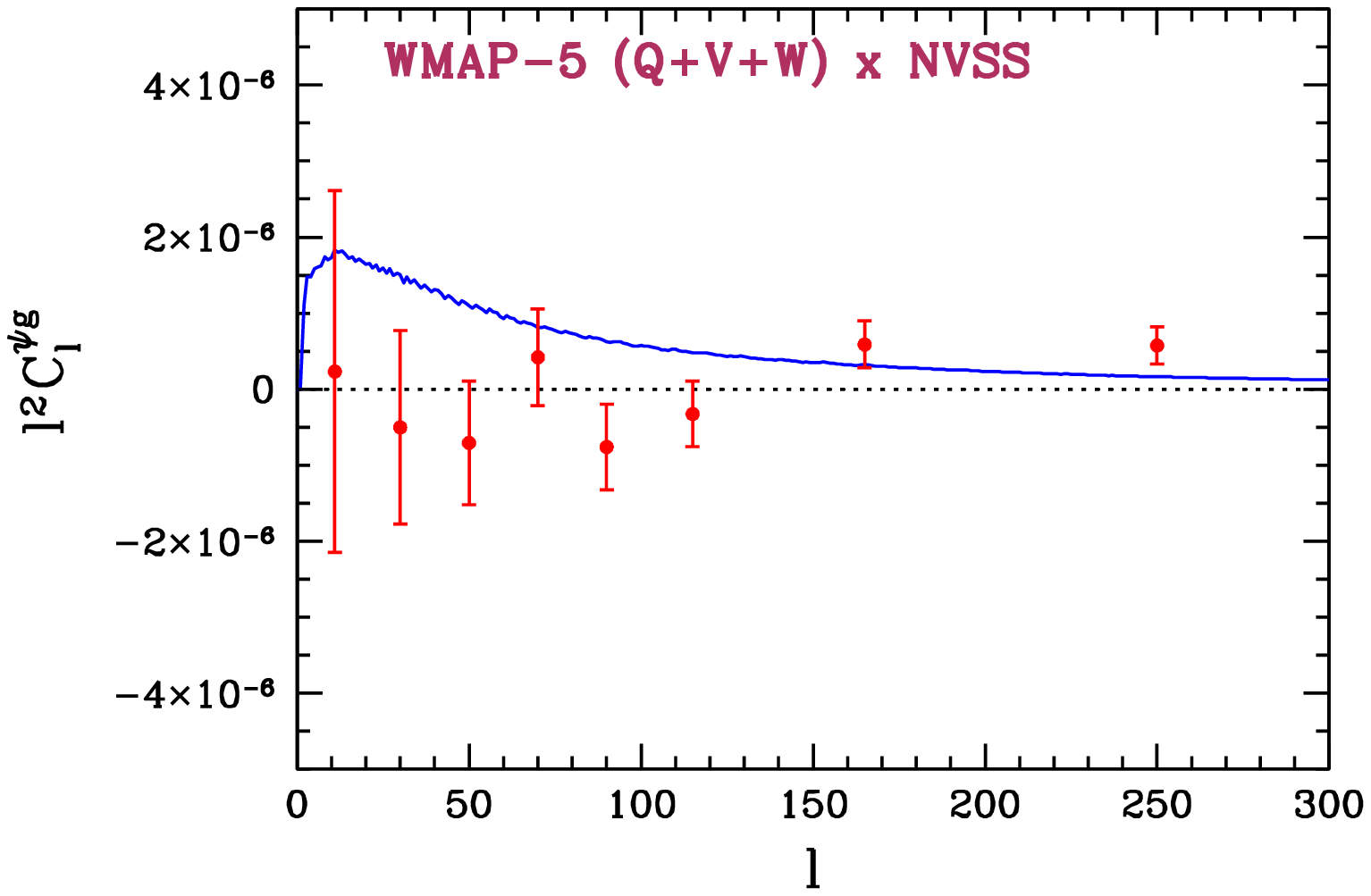}\qquad
\includegraphics[bb=50 195 465 480,width=8.5cm]{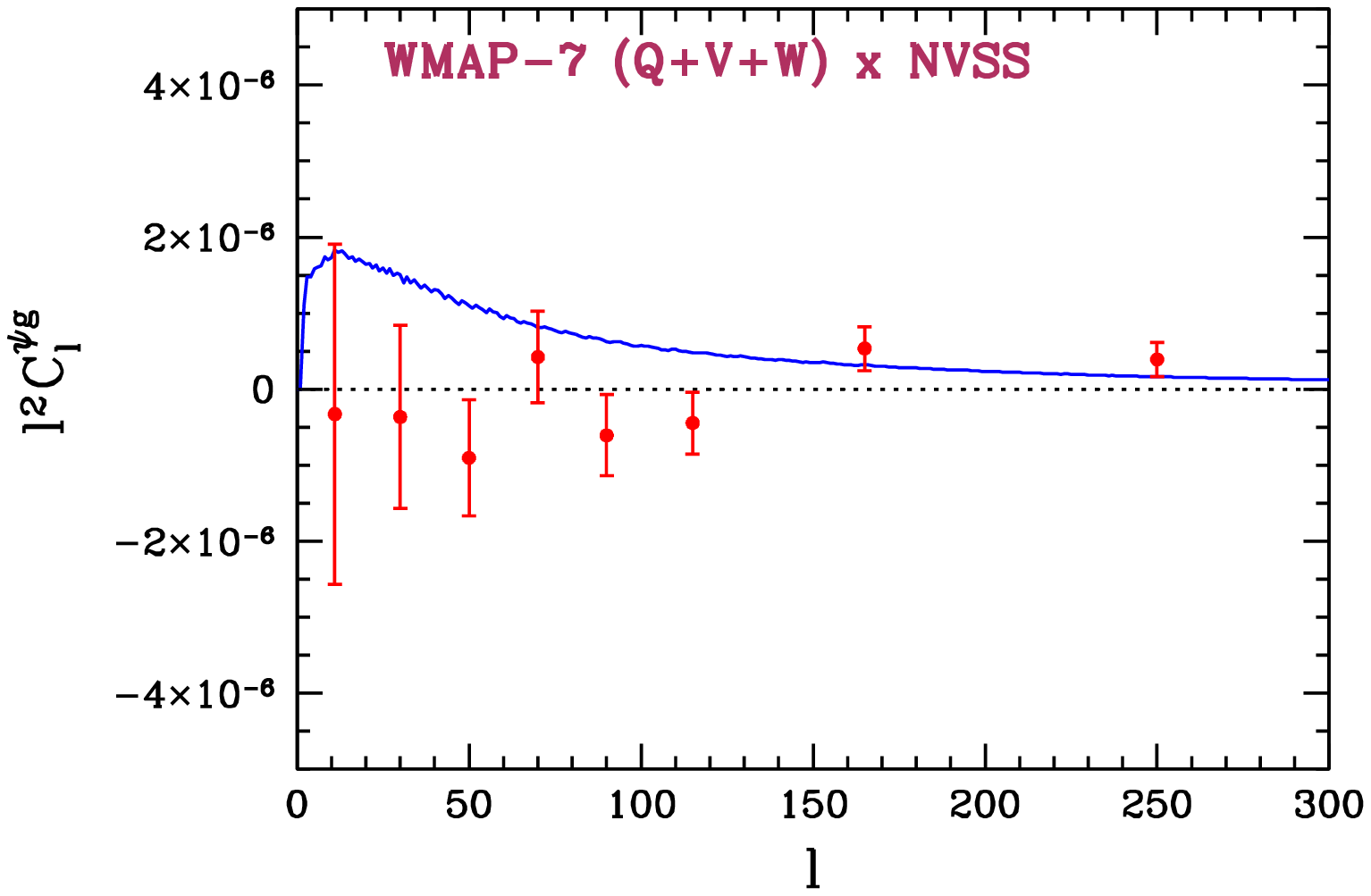}
\caption{(\pthree) The curl null tests for WMAP$\times$ NVSS are calculated from Eq. (\ref{curl-cross}). The Kp0 mask is used to remove the contaminated regions of the WMAP data. WMAP data are provided from two Q bands, two V bands and four W bands. The NVSS mask is applied to the galaxy map to remove bright sources and unobserved regions. The theoretical lensing-galaxy cross-power spectra with both WMAP and NVSS in galactic coordinates are shown in blue solid lines for comparison, and they are the same for all of the four panels. The curl amplitude for WMAP-1$\times$NVSS is $-0.11\pm0.47$, for WMAP-3$\times$NVSS is $0.00\pm0.35$, for WMAP-5$\times$NVSS is $0.05\pm0.31$, for WMAP-7$\times$NVSS is $-0.05\pm0.30$. As can be seen, all cross-power spectra for the curl null test are consistent with zero (black dotted line). }
\label{curlpower}
\end{figure*}
\begin{figure*}
\includegraphics[bb=50 195 465 480,width=8.5cm]{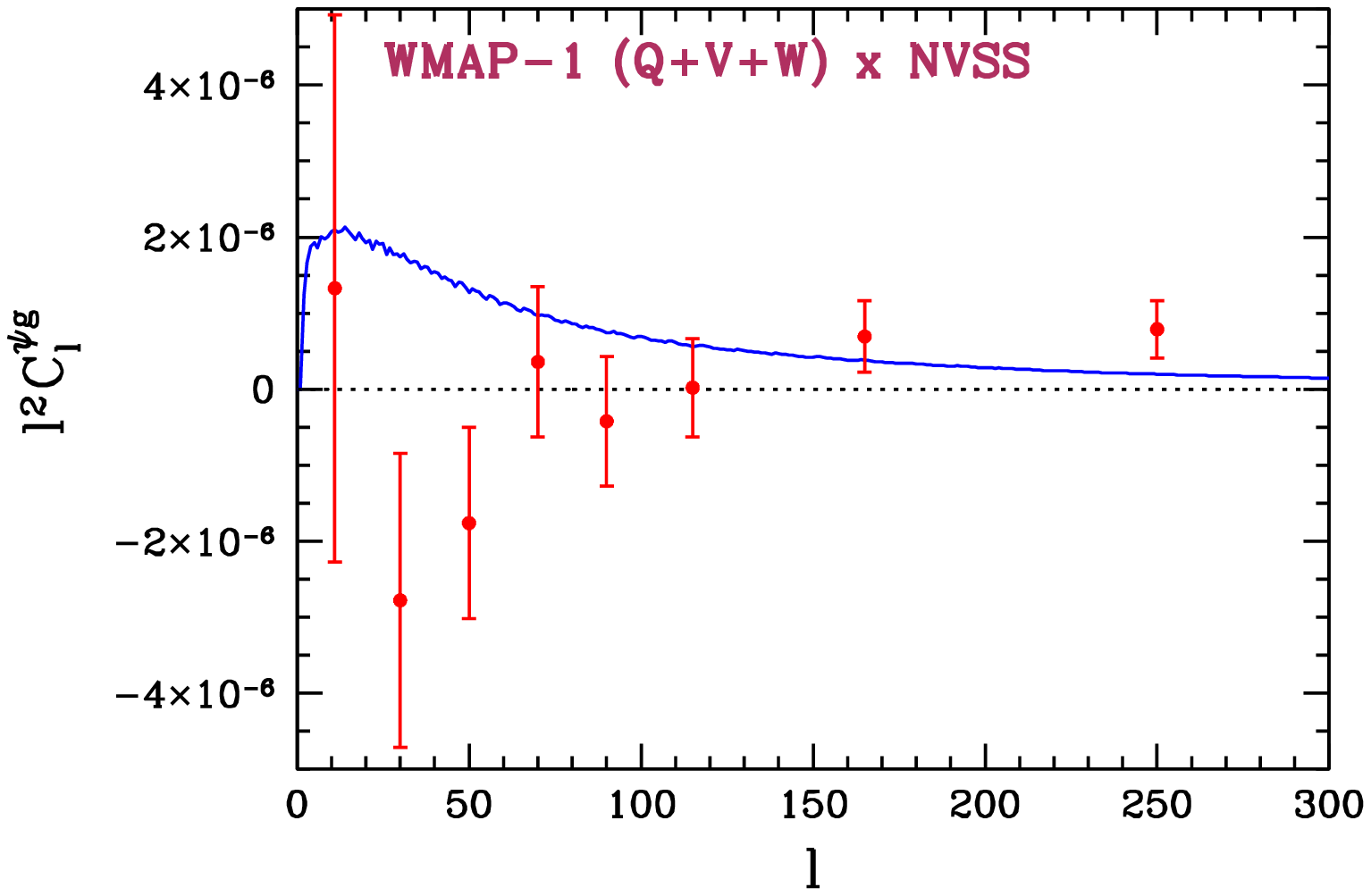}\qquad
\includegraphics[bb=50 195 465 480,width=8.5cm]{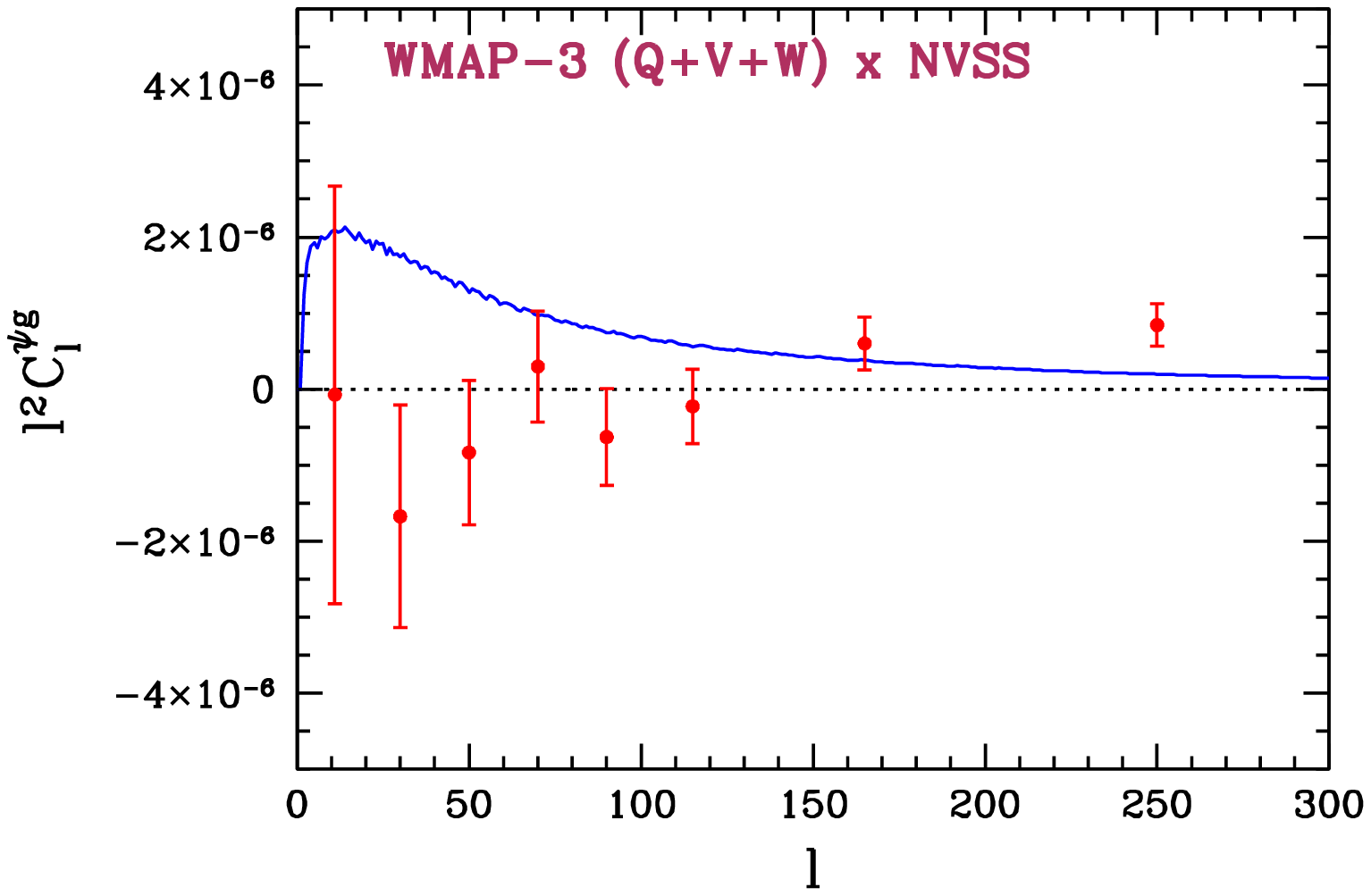} \\
\vspace{0.5cm}
\includegraphics[bb=50 195 465 480,width=8.5cm]{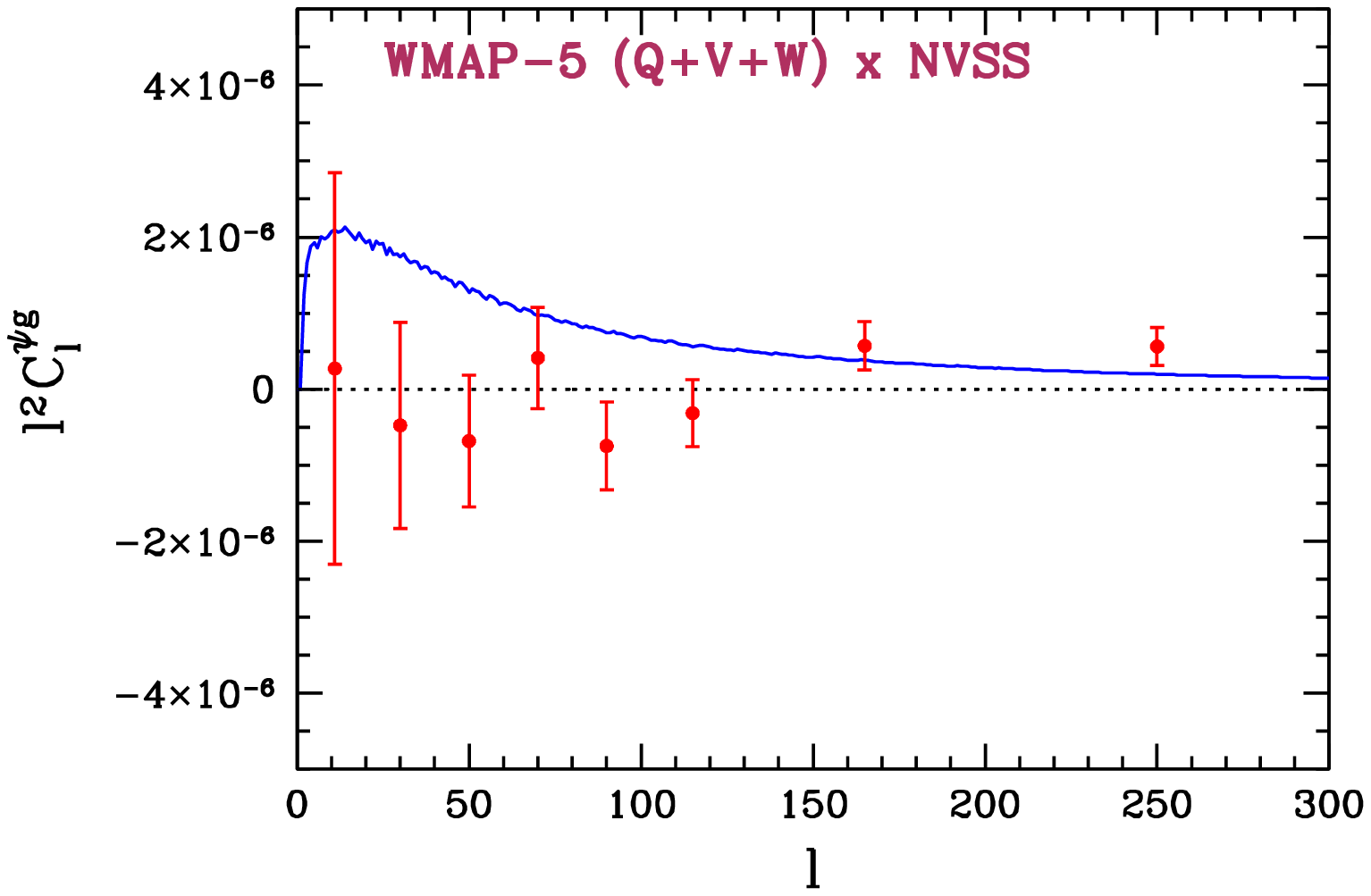}\qquad
\includegraphics[bb=50 195 465 480,width=8.5cm]{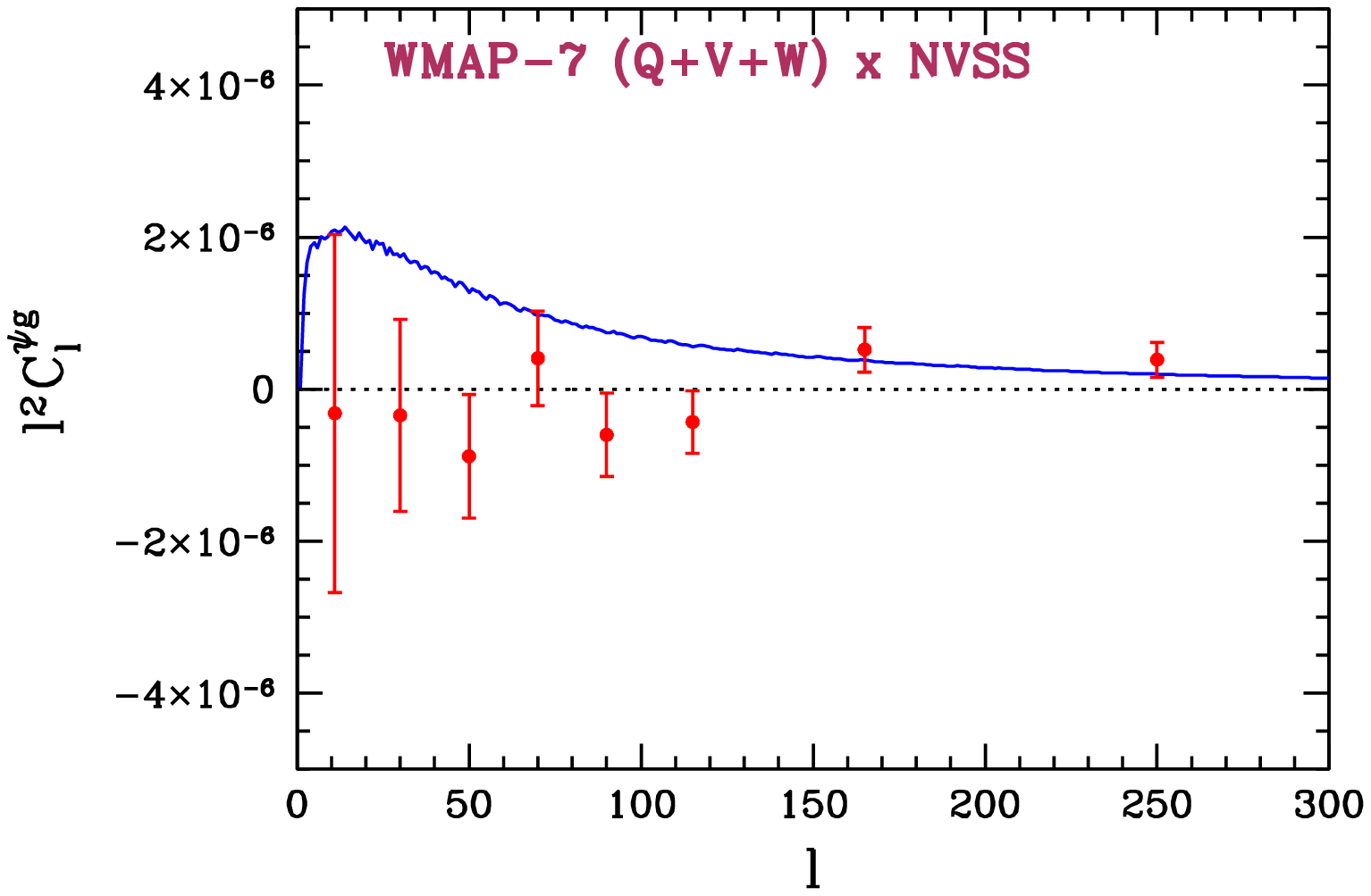}
\caption{(\pseven) The curl null tests for WMAP$\times$ NVSS are calculated from Eq. (\ref{curl-cross}). See Fig.~\ref{curlpower} for detailed descriptions. The curl amplitude for WMAP-1$\times$NVSS is $-0.03\pm0.41$, for WMAP-3$\times$NVSS is $0.04\pm0.31$, for WMAP-5$\times$NVSS is $0.07\pm0.28$, for WMAP-7$\times$NVSS is $-0.03\pm0.26$. As can be seen, all cross-power spectra for the curl null test are consistent with zero (black dotted line).}
\label{curlpowerpara2}
\end{figure*}

\section{Forecast for Future Experiments}
\label{sec:forcast}

\begin{figure}
\includegraphics[bb=50 195 465 480,width=8.5cm]{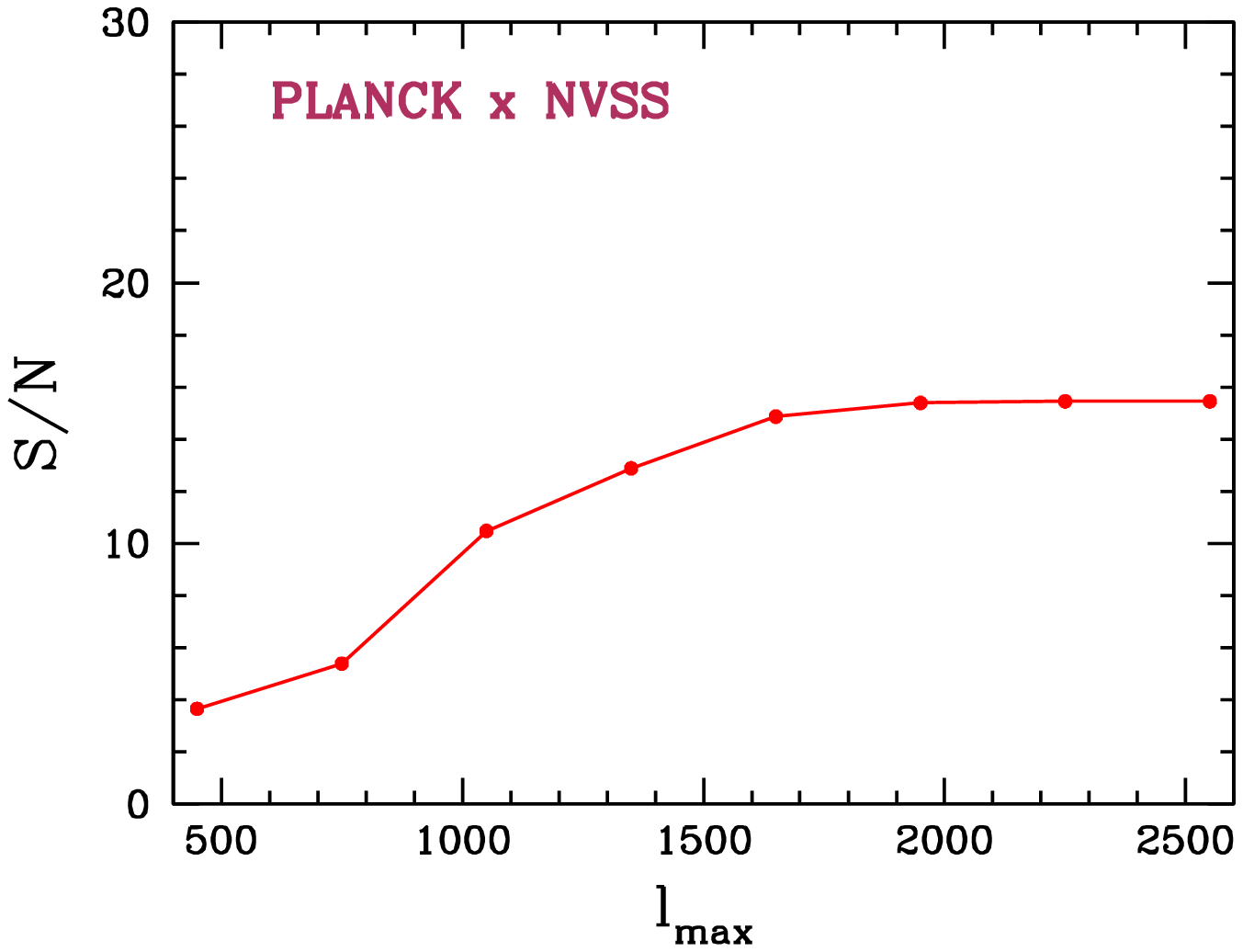}
\caption{The signal-to-noise ratio for the lensing-galaxy cross correlation between Planck and NVSS as a function of the maximum multipole used in the analysis.} \label{planck-nvss}
\end{figure}
The revealed cross correlation between WMAP and NVSS hints that the detection significance would be further enhanced if the precision of the CMB data were improved. The upcoming Planck data will improve upon WMAP, so we expect that the cross correlation between Planck and NVSS will be more significant. To predict the optimal bound on the detection signal-to-noise ratio for lensing-galaxy cross correlation we first calculate the equivalent noise $N_l^{\phi g}$ from the following equation
\begin{equation}
N_l^{\phi g}=\left[N_l^{\phi\phi}N_l^{gg}\right]^{1/2}.\label{crossNoise}
\end{equation}
where $N_l^{\phi\phi}$ is the lensing reconstruction noise~\cite{Okamoto:2003zw} and $N_l^{gg}$ is the galaxy shot-noise. The efficient algorithm for calculating $N_l^{\phi\phi}$ is given in Refs.~\cite{Smith:2010gu,Feng:2011jx}. This reconstruction noise can be minimized by combining different CMB channels and the minimum noise is
\begin{equation}
N_l^{\rm min, \phi\phi}=\frac{1}{\sum_{\nu}\left[N_l^{\nu,
\phi\phi}\right]^{-1}},
\end{equation}
Both of the noise spectra effectively propagate the uncertainty $\Delta C_l^{\phi g}$ into the cross-power spectrum $C_l^{\phi g}$. Specifically, we express it as
\begin{equation}
\Delta C_l^{\phi g}=\sqrt{\frac{2}{(2l+1)f_{\rm sky}}}\ (C_l^{\phi
g}+N_l^{\phi g}).\label{fisherErrorbar}
\end{equation}
The optimal bound is then determined from
\[
\left[ \sum_l \left(\frac{C_l^{\phi g}}{\Delta C_l^{\phi
g}}\right)^2\right]^{1/2}\,.
\]

The redshift distribution Eq.~(\ref{redshiftdistri}) was used and the galaxy bias was set equal to $b_g=1$. The instrumental properties for Planck are given in Refs.~\cite{Miller:2008zi,Planck:2006aa}. We show the signal-to-noise ratio for Planck with NVSS as a function of $l_{\text{max}}$ in Fig.~\ref{planck-nvss}. We find that the highest signal-to-noise ratio, i.e.\ $15\sigma$, saturates at $l_{\text{max}}=2000$. Since the lensing-galaxy cross-power spectrum scales as $C_l^{\phi g}\propto b_g$ as illustrated by Eq. (\ref{clphig}), the amplitude of this cross-power spectra is degenerate with the galaxy bias and the signal-to-noise for the cross-power spectrum can also serve as a prediction of the detection significance for the galaxy bias. Thus, we see Planck can detect $b_g$ with high precision which will lead to a better understanding of the correlation between the baryonic matter distribution and the dark matter distribution.

\section{Conclusion}
\label{sec:con}

We have calculated the lensing-galaxy cross-power spectra using WMAP and NVSS and the full covariance matrix to filter the data sets. Specifically, we performed a thorough analysis of WMAP-1, WMAP-3, WMAP-5 and WMAP-7 raw DAs. The cross correlations between WMAP-5, -7's 8 DAs (2Q-bands$+$2V-bands$+$4W-bands) with NVSS clearly and firmly show signals at $>3\sigma$ level. We took the effects of gradient stripes into account for the NVSS data, and determined the significance without and with gradient stripes removed. The major effects caused by the stripes can be seen from the first bin of either the galaxy auto-power spectrum or the lensing-galaxy cross-power spectrum; the first bin decreases if the gradient stripes are marginalized over. However, gradient stripes do not affect the lensing-galaxy correlation (compare Refs.~\cite{Schiavon:2012fc,Nolta:2003uy,MNR:MNR9764}). We have explicitly shown all these results in Tables \ref{lensingtable1} and \ref{lensingtable2}. In these two tables, column $\mathcal{C}^a$ are the results without the gradient stripes removed and column $\mathcal{C}^b$ are our main results with the gradient stripes removed. In order to validate the lensing-galaxy cross correlations, we produced a NVSS galaxy map in equatorial coordinates directly from the NVSS catalog and cross correlated it with the WMAP DA which is in galactic coordinates, we find that all the lensing-galaxy cross correlation amplitudes are negligible. 

We investigated the impact of different NVSS pixelization resolutions and found no effect. We compared the sensitivities of the estimators both with the full and diagonal covariance matrix and found that the former more effectively reduces the variance, which is mainly caused by the sky-cut and the inhomogeneous instrumental noise. However, the former scheme involves the inversion of a large matrix which is computationally challenging.

We predicted the detection significance for the lensing-galaxy cross correlation or the galaxy bias for the upcoming Planck data with NVSS and found the detection significance will be improved by a factor of 5. 

The minimum variance of the estimator assumes that the CMB and galaxy overdensity modes are Gaussian. However, if the CMB contains gravitational lensing, the bispectrum $\langle TTg\rangle$ is not zero; it induces an additional variance as indicated in Eq.~(\ref{NG}). We analytically and numerically confirm that this variance is actually not noticeable for WMAP and NVSS as pointed out in Ref.~\cite{Smith:2011we}. Furthermore, being aware of the potential non-Gaussian shape of the probability distribution function (PDF)~\cite{Smith:2012ty}, we specifically investigate the PDF of the cross-power spectrum amplitude $\mathcal {C}$ in terms of Kolmogorov-Smirnov test, the skewness and the kurtosis. The diagnostic tests are shown in Table~\ref{kstable}. All the PDFs pass the Kolmogorov-Smirnov tests. All the PDFs are consistent with being Gaussian-distributed (Figs.~\ref{like1} and \ref{like2}).

The lensing-galaxy cross correlations effectively link the early universe to the late universe and the CMB is served as a back light casting the dark cosmic web (which is formed by the dark matter) throughout the major expansion history of the universe. The gravitational lensing is a powerful tool to decode the information of dark matter distribution from the CMB and the lensing-galaxy cross-correlations further unveil the relationship between baryonic matter and dark matter. 

\begin{figure*}
\includegraphics[bb=50 195 465 480,width=8.5cm]{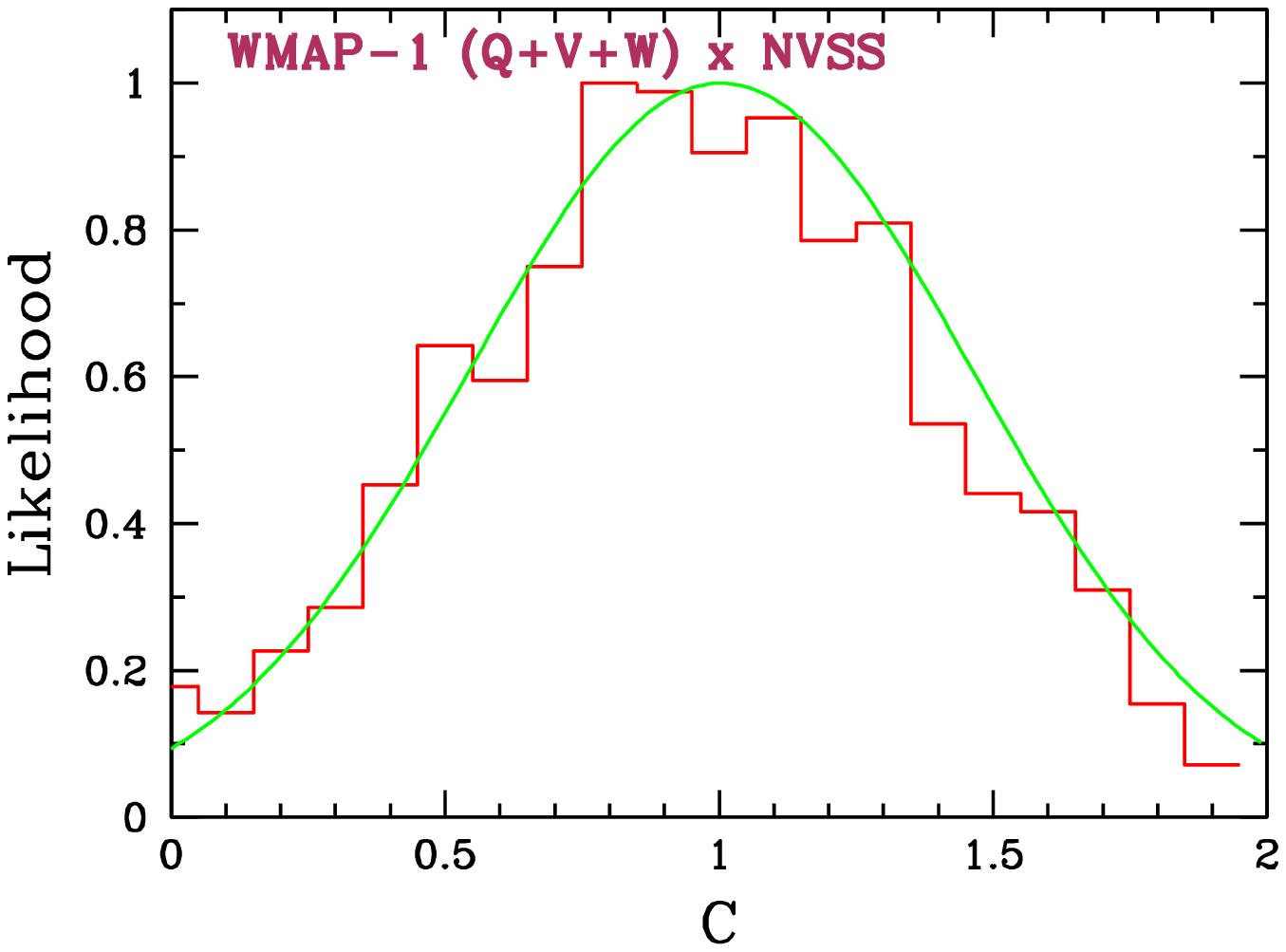}
\includegraphics[bb=50 195 465 480,width=8.5cm]{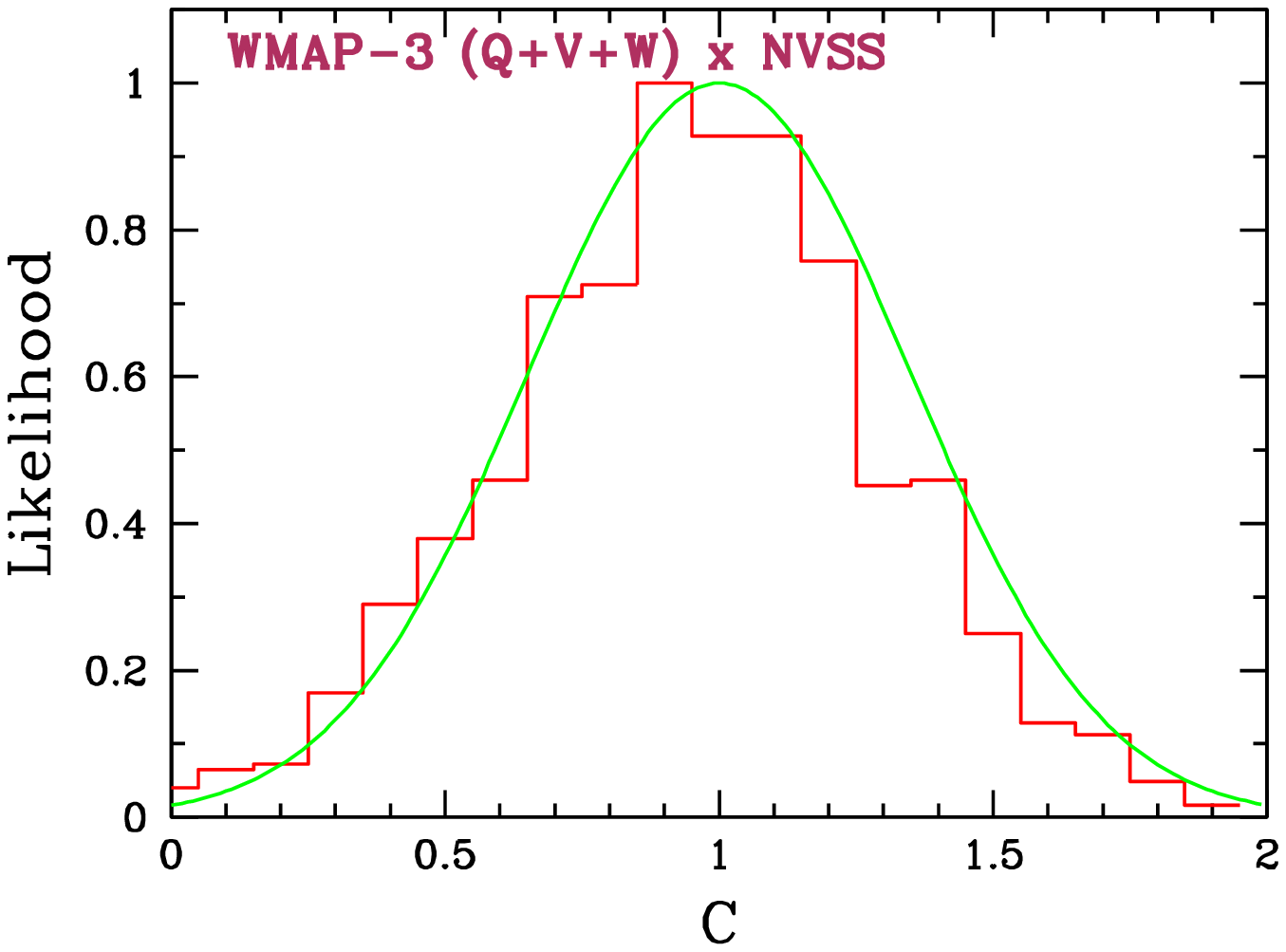}
\includegraphics[bb=50 195 465 480,width=8.5cm]{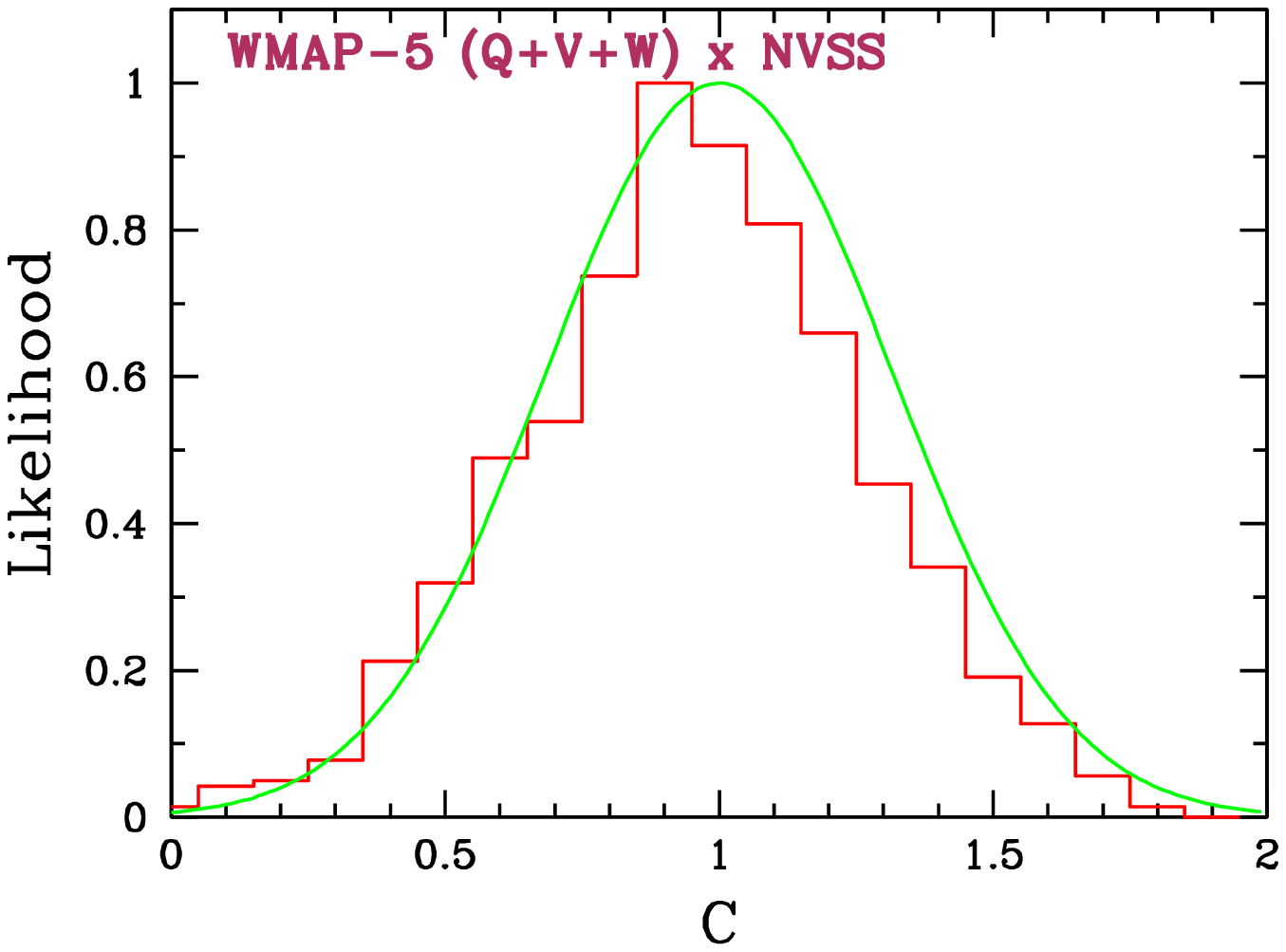}
\includegraphics[bb=50 195 465 480,width=8.5cm]{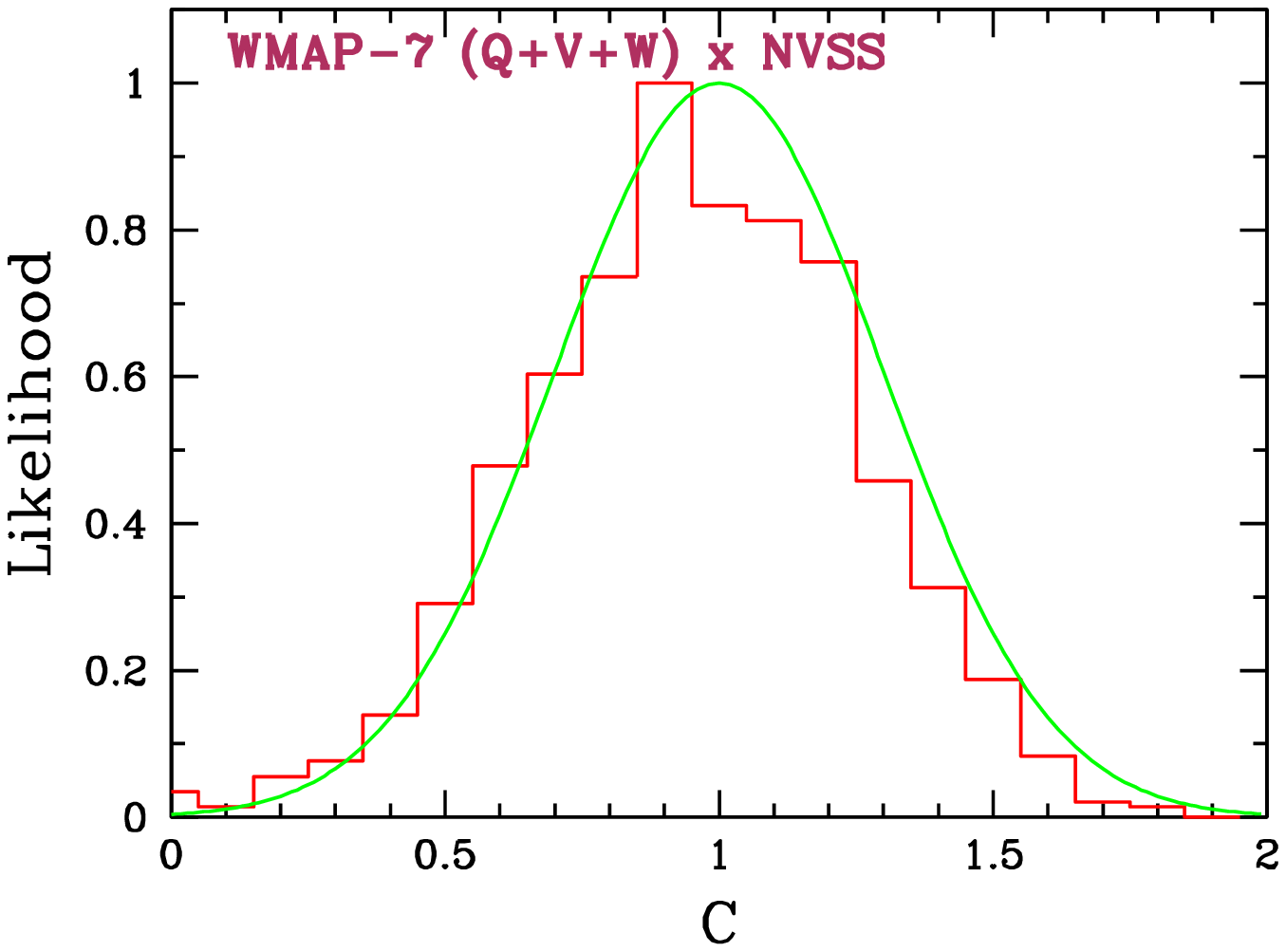}
\caption{(\pthree) Probability distribution function
for the lensing-galaxy cross correlation. The likelihood functions are normalized to 1. From 1000 simulations, a set of $\{\mathcal{C}\}$ is generated for each one of the subfigures, then by counting the frequency of $\mathcal{C}$ within a bin, a step-like function (red) is plotted. For comparison, Gaussian likelihood (green) is plotted using the mean and the variance of the set $\{\mathcal{C}\}$. } \label{like1}
\end{figure*}
\begin{figure*}
\includegraphics[bb=50 195 465 480,width=8.5cm]{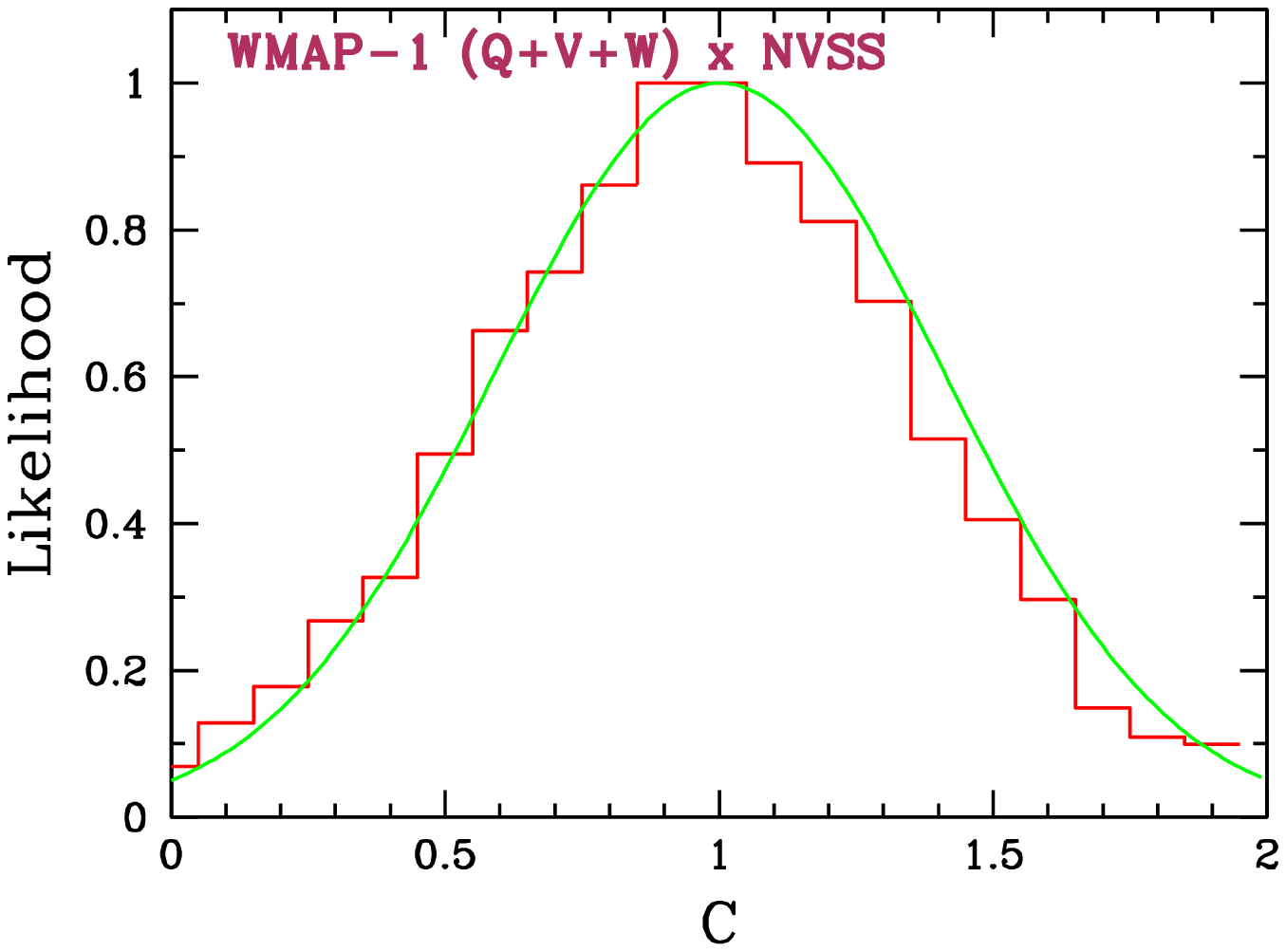}
\includegraphics[bb=50 195 465 480,width=8.5cm]{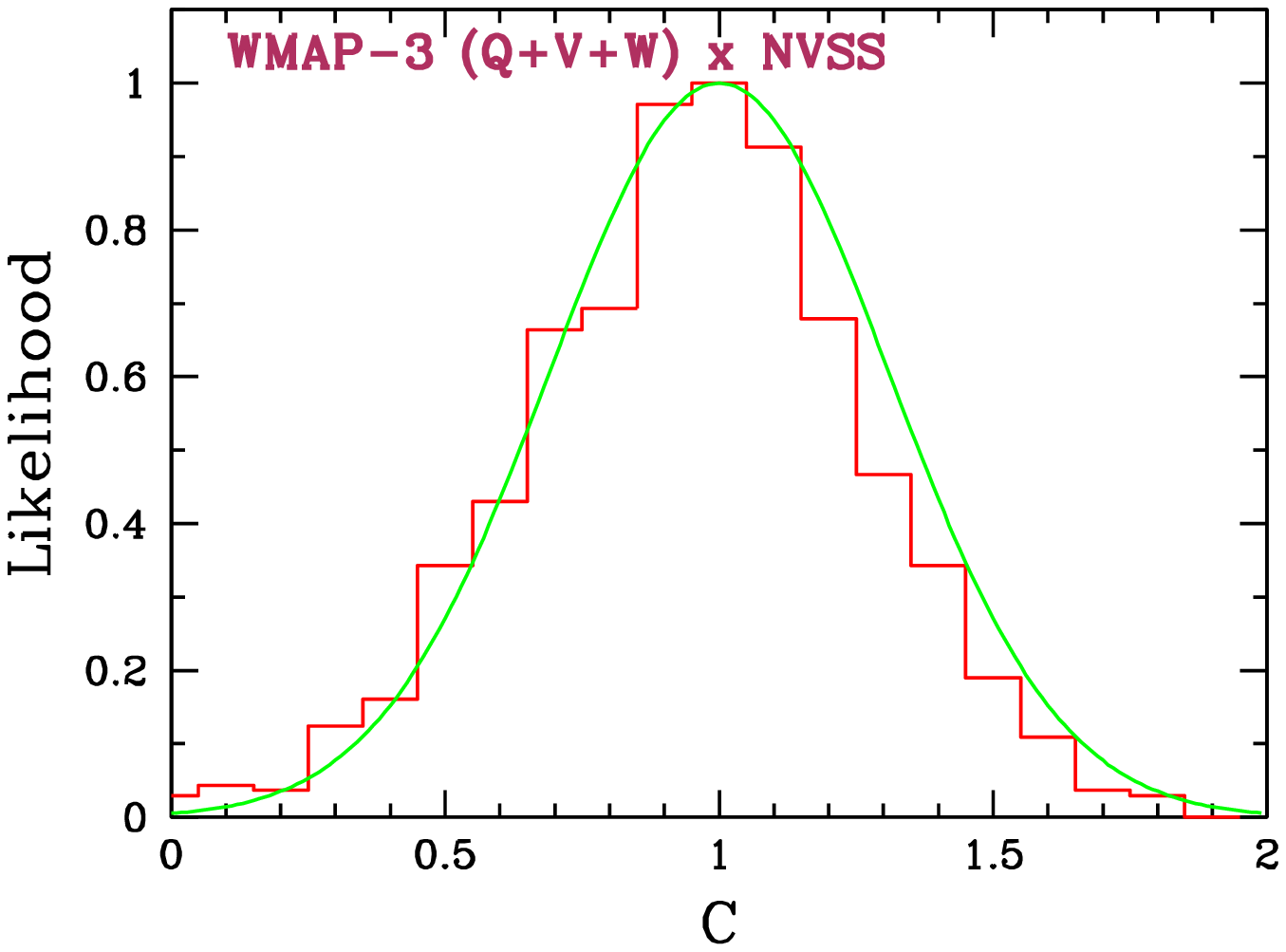}
\includegraphics[bb=50 195 465 480,width=8.5cm]{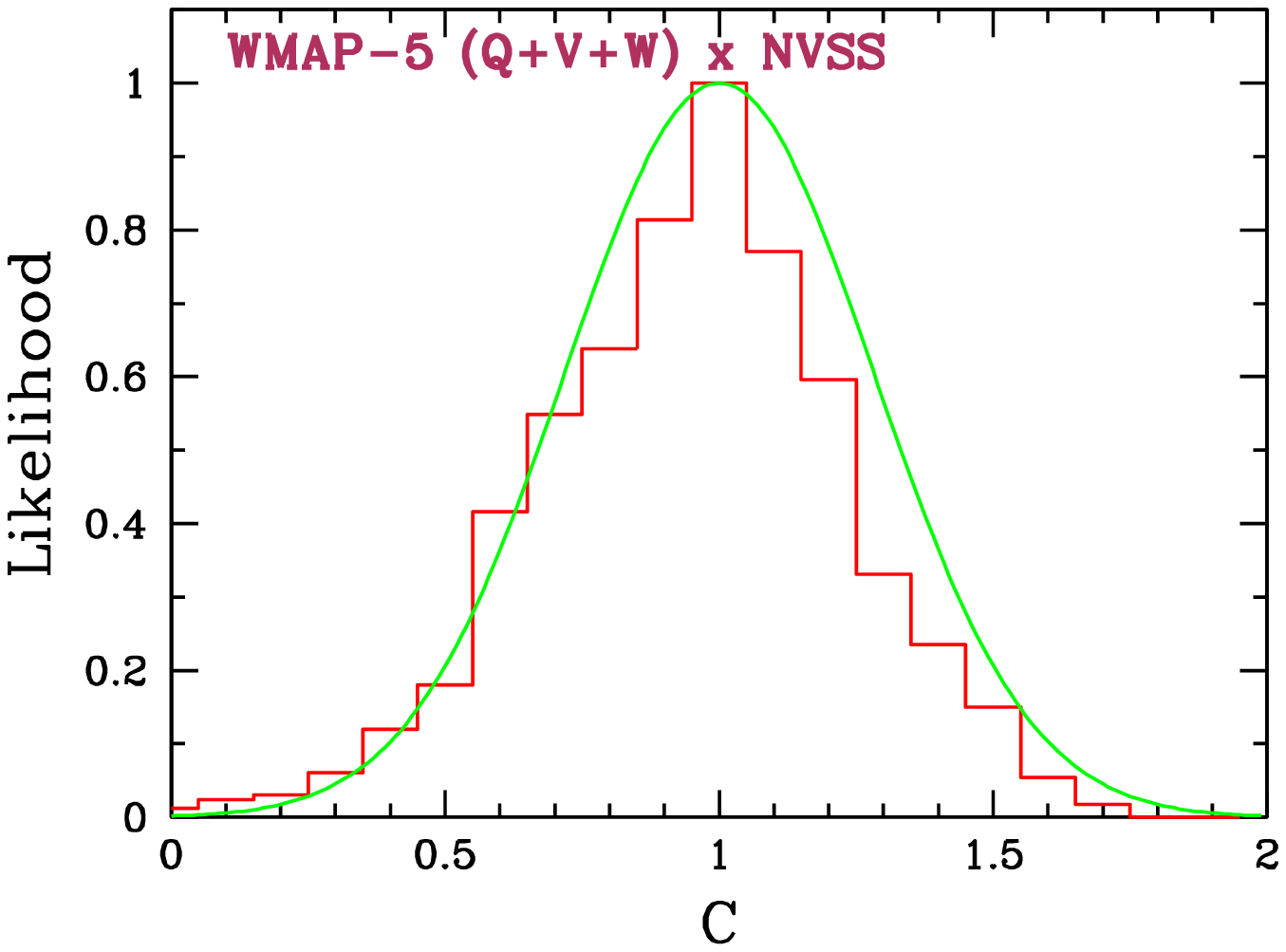}
\includegraphics[bb=50 195 465 480,width=8.5cm]{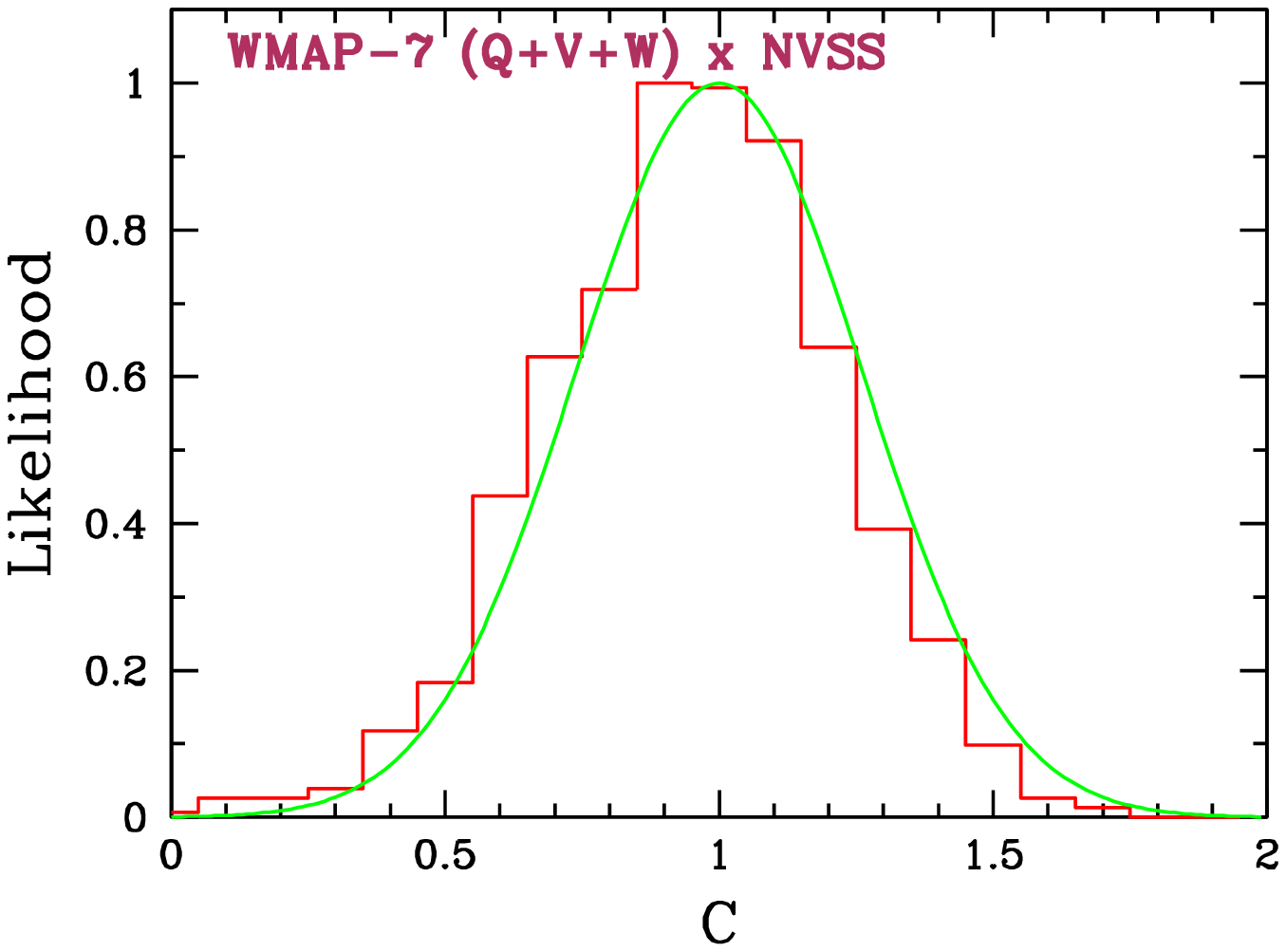}
\caption{(\pseven) Probability distribution function
for the lensing-galaxy cross correlation. See Fig.~\ref{like1} for detailed descriptions. } \label{like2}
\end{figure*}

\begin{acknowledgments}
We would like to acknowledge helpful discussions with Sudeep Das, Duncan Hanson, Christian Reichardt, Meir Shimon, and Amit Yadav. We acknowledge the use of \camb, \healpix\footnotemark[1]~\cite{healpixpaper}, and \lapack\ software packages and the LAMBDA archive. The computational resources required for this work were accessed via the GlideinWMS~\cite{Sfiligoi2009} on the Open Science Grid~\cite{Pordes2008}. We are indebted to Frank Wuerthwein, Igor Sfiligoi, Terrence Martin, and Robert Konecny for their insight and support.
\end{acknowledgments}

\footnotetext[1]{\url{http://healpix.jpl.nasa.gov/}}

\appendix
\section{MULTIGRID-PRECONDITIONED COMPLEX CONJUGATE GRADIENT INVERSION}
\label{app:mpcg}

Given the signal covariance matrix ${\bf S}$ and the noise covariance matrix ${\bf N}$, and an array of the CMB modes ${\bf a}$, we define another covariance matrix $\mathbf{A}=\mathbf{I}+\mathbf{S}^{1/2}\mathbf{N}^{-1}\mathbf{S}^{1/2}$, and two vectors $\mathbf{x}=\mathbf{S}^{1/2}\mathbf{a}$, and $\mathbf{y}=\mathbf{S}^{1/2}\mathbf{N}^{-1}\mathbf{a}$. For the problem $\bf A\bf x=\bf y$, we write down the equations for constructing the matrix ${\bf A}$ and the vector ${\bf y}$,
\begin{eqnarray}
N^{-1}_{lml'm'}&=& \sum_{\nu}p_lb^{(\nu)}_lp_{l'}b^{(\nu)}_{l'}\nonumber\\
&&\times\int d\textbf{n}\ Y^{\ast}_{lm}(\textbf{n})Y_{l'm'}(\textbf{n})\left[\frac{\rm
M({\bf n})}{\sigma^{2}}\right]^{(\nu)}\,, \label{coaddedN} 
\end{eqnarray}
\begin{eqnarray}
{}[N^{-1}a]_{lm} &=& \sum_{\nu}p_lb^{(\nu)}_l\nonumber\\&&\times\int
d\textbf{n}\ Y^{\ast}_{lm}(\textbf{n})\left[\frac{\rm M({\bf n})\rm H({\bf
n})}{\sigma^{2}}\right]^{(\nu)}\,,\label{coaddedT} 
\end{eqnarray}
\begin{eqnarray}
w^{(\nu)}_{lm} &=& \int d\mathbf{n}\ Y_{lm}(\mathbf{n})\left[\frac{\rm M({\bf
n})}{\sigma^2}\right]^{(\nu)} \,.\label{invnoisemode}
\end{eqnarray}
In the above equations, $p_l$ is the window transfer function, $b_l^{(\nu)}$ is the specific beam transfer function corresponding to the DA of WMAP, and $\rm M({\bf n})$ is the mask map. For WMAP, $\nu=Q_1, Q_2, V_1, V_2, W_1, W_2, W_3, W_4$, $\rm H({\bf n})=T(\mathbf{n})$ and $\rm M({\bf n})$ is the Kp0 mask. For NVSS, $\nu=1$ and $\rm H({\bf n})=g(\mathbf{n})$. Since NVSS has $45$ arc-second FWHM resolution~\cite{Condon:1998iy}, we set $b_l^{(1)} =1$ as NVSS's beam transfer function. The correspondence between the continuum and discrete forms of  integration on the sphere is,
\begin{equation}
\int d\mathbf{n} \rightarrow
\frac{4\pi}{N_{\textrm{pix}}}\displaystyle\sum_{j},
\end{equation}
where $j$ denotes the pixel index according to the \healpix\ pixelization scheme and $N_{\textrm{pix}}$ is the total number of  pixels.

For comparison, we also use a suboptimal estimator which only takes the diagonal elements of the inverse noise matrix $N^{-1}_{lml'm'}$ shown in Eq.~(\ref{coaddedN}).

The filtering using the covariance matrix requires us to solve the linear equation $\mathbf{A}\mathbf{x}=\mathbf{y}$. We use the preconditioned conjugate gradient iteration to solve it, and the initial condition is chosen to be
\begin{eqnarray}
{\bf x}^{(0)}&=&0,\nonumber\\
{\bf r}^{(0)}&=&{\bf y},\nonumber\\
{\bf p}^{(1)}&=&\tilde{{\bf A}}^{-1}{\bf y},
\end{eqnarray}
with the preconditioner defined as
\begin{equation}
\tilde{{\bf A}}^{-1}=\left(
\begin{array}{cc}
{\bf A}_0^{-1}& {\bf 0}\\
{\bf 0}&{\bf A}^{-1}_{\Delta}
\end{array} \right),
\end{equation}
here ${\bf A}_{\Delta}$ is the diagonal element of the matrix ${\bf A}$.

The iteration procedure is~\cite{Hirata:2004rp}
\begin{eqnarray}
\mathbf{x}^{(i)}&=&\mathbf{x}^{(i-1)}+\frac{\mathbf{r}^{(i-1)}\tilde{\mathbf{A}}^{-1}
\mathbf{r}^{(i-1)}}{\mathbf{p}^{(i)}\mathbf{A}\mathbf{p}^{(i)}}\mathbf{p}^{(i)},\nonumber\\
\mathbf{r}^{(i)}&=&\mathbf{y}-\mathbf{A}\mathbf{x}^{(i)},\nonumber\\
\mathbf{p}^{(i)}&=&\tilde{\mathbf{A}}^{-1}\mathbf{r}^{(i-1)}+\frac{\mathbf{r}^{(i-1)}\tilde{\mathbf{A}}^{-1}\mathbf{}r^{(i-1)}}{\mathbf{r}^{(i-2)}
\tilde{\mathbf{A}}^{-1}\mathbf{r}^{(i-2)}}\mathbf{p}^{(i-1)}\label{pcgiter}.
\end{eqnarray}

From Eq. (\ref{pcgiter}), we find that two operations $\tilde {\bf A}^{-1}{\bf r}$ and ${\bf A}{\bf p}$ are computationally demanding if we evaluate them directly because ${\bf A}$ and ${\bf A}_0$ are $10^6\times10^6$ matrix.

In order to achieve the necessary efficiency, we recursively precondition the matrix ${\bf A}$ on a much coarser grid. The preconditioner is
\begin{equation}
\tilde{{\bf A}}^{-1}=\left(
\begin{array}{cc}
\tilde {\bf A}_0^{-1}& {\bf 0}\\
{\bf 0}&{\bf A}^{-1}_{\Delta}
\end{array} \right),
\end{equation}
and on the coarser grid the preconditioner is $\tilde {\bf A}_0^{-1}$. This multigrid strategy enables us to directly store the matrix $\tilde {\bf A}_0$ on the coarsest grid and we can further analytically express the smallest inversion problem as follows\footnote[2]{In the following, we denote subscripts $l_i$ or $l_i,m_i$ by $i$ for simplicity, so that  $p_{l_i} \to p_i $, $Y_{l_im_i} \rightarrow  Y_i$,  $N^{-1}_{l_1m_1l_2m_2} \rightarrow N^{-1}_{12}$  etc. \label{foot1}}
\begin{eqnarray}
N_{12}^{-1}&=&\displaystyle\sum_{\nu}\int
p_1b^{(\nu)}_1Y_1^{\ast}p_2b^{(\nu)}_2Y_2\displaystyle\sum_{3}w^{(\nu)}_3Y_3\nonumber\\
&=&\displaystyle\sum_{\nu}\displaystyle\sum_{3}w^{(\nu)}_3\sqrt{\frac{(2l_1+1)(2l_2+1)(2l_3+1)}{4\pi}}\nonumber\\
&\times&(-1)^{m_1}\left(
\begin{array}{ccc}
l_1& l_2&l_3\\
0&0&0
\end{array} \right)\left(
\begin{array}{ccc}
l_1& l_2&l_3\\
-m_1&m_2&m_3
\end{array} \right)\nonumber\\&&p_1b^{(\nu)}_1p_2b^{(\nu)}_2,\label{analyticinvN}
\end{eqnarray}
then the problem of preconditioning $\bf A$ with $\tilde {\bf A}_0$ on the finer grid can be iteratively solved by using Eq.~(\ref{pcgiter}). For this work we use three levels of the grids: (1) $N_{\rm side}=512, l_{\rm max}=1000$, (2) $N_{\rm side}=256, l_{\rm max}=400$, (3) $N_{\rm side}=128, l_{\rm max}=200$. We split the covariance matrix on the third grid at $l_{\rm split}=30$ to construct the minimum inversion problem.

For the coarsest grid, we explicitly calculate the inverse noise matrix $N_{12}^{-1}$ [Eq. (\ref{analyticinvN})] using \lapack. The iteration process also needs the multiplication for $\mathbf{A}\mathbf{\lambda}$, and this can be  computed efficiently by doing spherical harmonic transformations:
\begin{eqnarray}
\mathbf{A}\mathbf{\lambda}&=&\displaystyle\sum_4(I+S^{1/2}N^{-1}S^{1/2})_{14}\lambda_4\nonumber\\
&=&\lambda_1+\sum_{\nu}p_1b^{(\nu)}_1S^{1/2}_1\biggl[\int d {\bf n}\
Y_1^{\ast}\left[\frac{\rm M({\bf
n})}{\sigma^2}\right]^{(\nu)}\nonumber\\&& \times \left(\sum_4p_4b^{(\nu)}_4S^{1/2}_4\lambda_4Y_4\right)\biggr].
\end{eqnarray}

\section{NON-GAUSSIANITY}
\label{app:ng}

There are several possible non-Gaussian effects generated by using a nonzero bispectrum in the simulation. These can potentially bias our results. We analytically calculate this non-Gaussian bias in this appendix.

We define the bispectrum by
\begin{equation}
\langle a_1a_2g_3\rangle=b_{123}G(123),
\end{equation}
where $b_{123}=(f_{123}C_{2}^{\rm TT}+f_{213}C_{1}^{\rm TT})C_{l_3}^{\phi g}$, (see footnote 2 for notation)
and
\begin{eqnarray}G(123)&=&\sqrt{\frac{(2l_1+1)(2l_2+1)(2l_3+1)}{4\pi}}\left(
\begin{array}{ccc}
l_1& l_2&l_3\\
0&0&0
\end{array} \right)\nonumber\\&&\left(
\begin{array}{ccc}
l_1& l_2&l_3\\
m_1&m_2&m_3
\end{array} \right).
\end{eqnarray}

 The estimator is
\begin{equation}
\hat C=\frac{1}{\mathcal {F}}(\hat C_A-\hat C_B)
\end{equation}
where
\begin{equation}
\hat C_A=\frac{1}{2}\displaystyle\sum_{123}b_{123}G(123)\tilde
a_1\tilde a_2\tilde g_3
\end{equation}
and
\begin{equation}
\hat C_B=\frac{1}{2}\displaystyle\sum_{123}b_{123}G(123)\left[C^{\rm TT}\right]^{-1}_{12}\tilde g_3.
\end{equation}

We define 
\begin{eqnarray*}
\tilde a &=& C^{-1}a, \nonumber \\
f_k &=& \frac{1}{2}\sum b_{12k}G(12k)\left[C^{\rm TT}\right]^{-1}_{12},\nonumber \\
\langle \tilde a_1\tilde a_2\rangle &=& \left[C^{\rm TT}\right]^{-1}_{12},\nonumber \\
\langle \tilde g_1\tilde g_2\rangle &=& \left[C^{\rm gg}\right]^{-1}_{12}\,.
\end{eqnarray*}
The summation index $i$ denotes  a sum over $l_im_i$.

We define the normalization as
\begin{eqnarray}
\mathcal
{F}&=&\frac{1}{2}\displaystyle\sum_{123456}b_{123}b_{456}G(123)G(456)\nonumber\\
&\times&\left[C^{\rm TT}\right]^{-1}_{14}\left[C^{\rm TT}\right]^{-1}_{25}\left[C^{\rm gg}\right]^{-1}_{36}.
\end{eqnarray}

The variance of the estimator $\hat C$ is $\sigma^2(\hat C)$ which has contributions from three parts,
\begin{equation}
\sigma^2(\hat C)=\sigma^2(\hat C_A)+\sigma^2(\hat
C_B)-2\sigma^2(\hat C_A\hat C_B).
\end{equation}
Now we explicitly determine the three variances. For the second term, we have the relation
\begin{equation}
\langle\tilde a_1\tilde a_2\tilde
g_3\rangle=\displaystyle\sum_{1'2'3'}\left[C^{\rm TT}\right]^{-1}_{11'}\left[C^{\rm TT}\right]^{-1}_{22'}\left[C^{\rm gg}\right]^{-1}_{33'}b_{1'2'3'}G(1'2'3'),
\end{equation}
so the last two variance terms can be easily expressed as
\begin{equation}
\sigma^2(\hat C_B)=\sigma^2(\hat C_A\hat C_B)=f^T\left[C^{\rm gg}\right]^{-1}f.
\end{equation}
For the first term, it is
\begin{widetext}
\begin{eqnarray}
\sigma^2(\hat C_A)
&=&\frac{1}{4}\displaystyle\sum_{123456}b_{123}b_{456}G(123)G(456)\langle\tilde a_1\tilde a_2\tilde g_3\tilde a_4\tilde
a_5\tilde g_6\rangle - \frac{1}{4}\displaystyle\sum_{123456}b_{123}b_{456}G(123)G(456) \langle\tilde a_1\tilde a_2\tilde g_3\rangle\langle\tilde
a_4\tilde a_5\tilde g_6\rangle,\nonumber \\
\end{eqnarray}
and can be expanded as
\begin{eqnarray}
&&\frac{1}{4}\displaystyle\sum_{123456}b_{123}b_{456}G(123)G(456)\langle\tilde
a_1\tilde a_2\tilde g_3\tilde a_4\tilde
a_5\tilde g_6\rangle\nonumber \\
&=&\frac{1}{4}\displaystyle\sum_{123456}b_{123}b_{456}G(123)G(456)\biggl\{\Bigl[
\underbrace{\langle\tilde a_1\tilde a_2\tilde
g_3\rangle\langle\tilde a_4\tilde a_5\tilde g_6\rangle}_\text{second
term} +\langle\tilde a_4\tilde a_5\tilde g_3\rangle\langle\tilde
a_1\tilde
a_2\tilde g_6\rangle
+\langle\tilde a_1\tilde a_4\tilde g_3\rangle\langle\tilde
a_2\tilde a_5\tilde g_6\rangle+\langle\tilde a_2\tilde a_5\tilde
g_3\rangle\langle\tilde a_1\tilde a_4\tilde g_6\rangle \nonumber \\
&&+\langle\tilde a_1\tilde a_5\tilde g_3\rangle\langle\tilde
a_2\tilde a_4\tilde g_6\rangle+\langle\tilde a_2\tilde a_4\tilde
g_3\rangle\langle\tilde a_1\tilde a_5\tilde
g_6\rangle\Big]_{3\cdot3}+ \Big[\underbrace{\langle\tilde a_1\tilde a_5\rangle\langle\tilde
a_2\tilde a_4\rangle\langle\tilde g_3\tilde g_6\rangle+\langle\tilde
a_1\tilde a_4\rangle\langle\tilde a_2\tilde
a_5\rangle\langle\tilde g_3\tilde g_6\rangle}_\text{normalization}+\langle\tilde a_1\tilde a_2\rangle\langle\tilde a_4\tilde
a_5\rangle\langle\tilde g_3\tilde
g_6\rangle\Bigr]_{2\cdot2\cdot2}\biggr\}.\nonumber \\
\end{eqnarray}
\begin{eqnarray}
\sigma^2(\hat C_A-\hat C_B)
&=&\mathcal {F}+\biggl\{\frac{1}{4}\displaystyle\sum_{123456}b_{123}b_{456}G(123)G(456)\Bigl[
\langle\tilde a_4\tilde a_5\tilde g_3\rangle\langle\tilde
a_1\tilde a_2\tilde g_6\rangle +\langle\tilde a_1\tilde a_4\tilde g_3\rangle\langle\tilde
a_2\tilde a_5\tilde g_6\rangle+\langle\tilde a_2\tilde a_5\tilde
g_3\rangle\langle\tilde a_1\tilde a_4\tilde g_6\rangle\nonumber\\
&&+\langle\tilde a_1\tilde a_5\tilde g_3\rangle\langle\tilde
a_2\tilde a_4\tilde g_6\rangle+\langle\tilde a_2\tilde a_4\tilde
g_3\rangle\langle\tilde a_1\tilde a_5\tilde g_6\rangle\Bigr]\biggr\}_{\rm nonzero{\ }bispectrum}\nonumber\\
&=&O(b^2)+O(b^4)\label{NG}
\end{eqnarray}
\end{widetext}
When all the pieces are put together, we find that the nonvanishing bispectrum induces an extra term which is $O(b^4)$. We conclude that for WMAP, this contribution is very small and we numerically verified that this is indeed the case.

\end{document}